\documentclass[11pt,a4paper,twoside]{article}
\usepackage{axodraw}
\usepackage{epsfig}
\usepackage[latin1]{inputenc}
\usepackage{graphicx,bbm}
\usepackage{amsmath,amsfonts,amssymb}
\usepackage{mathrsfs}
\usepackage[footnotesize]{caption}
\def\be{\begin{equation}}
\def\ee{\end{equation}}
\def\ba{\begin{array}}
\def\bacc{\begin{array} {cc}}
\def\ea{\end{array}}
\def\bea{\begin{eqnarray}}
\def\eea{\end{eqnarray}}
\def\bd{\begin{displaymath}}
\def\ed{\end{displaymath}}

\def\epn{\epsilon_\nu}
\def\epd{\epsilon_d}
\def\epu{\epsilon_u}

\def\Yf{{\mathcal{Y}}^f}

\def\Box{ {\,\lower 0.9pt\vbox{\hrule\hbox{\vrule height0.2cm \hskip 0.2cm
\vrule height 0.2cm }\hrule}\,}}

\newcommand{\ts}{\text{s}}
\newcommand{\tc}{\text{c}}

\def\<{\left\langle}
\def\>{\right\rangle}

\def\MG{M_{\text{GUT}}}
\def\ME{M_{\text{EW}}}
\def\Gt{\text{GUT}}
\def\Ew{\text{EW}}

\def\beq{\begin{equation}}
\def\eeq{\end{equation}}
\def\bea{\begin{eqnarray}}
\def\eea{\end{eqnarray}}

\newcommand{\eq}[1]{Eq.~(\ref{#1})}

\begin{document}

\begin{titlepage}

\vspace*{-15mm}
\begin{flushright}
{\small KIAS-P10014}\\
{\small IC/2010/021}\\
{\small MPP-2010-58}
\end{flushright}
\vspace*{3mm}

\vspace{2ex}

\begin{center}
{
\bf\LARGE
Bridging flavour violation and leptogenesis in $\mathbf{SU(3)}$ family models
}\\
[8mm]

{\bf Lorenzo Calibbi},\\
Max-Planck-Institut f\"ur Physik (Werner-Heisenberg-Institut), D-80805 M\"unchen, Germany\\
calibbi@mppmu.mpg.de\\

{\bf Eung Jin Chun} \\
Korea Institute for Advanced Study, 207-43 Cheongryangri-dong, Dongdaemun-gu, Seoul 130-012, Korea\\
ejchun@kias.re.kr\\

and\\
{\bf Liliana Velasco-Sevilla}\\
The Abdus Salam International Centre for Theoretical Physics, Strada Costiera 11, Trieste 34151, Italy\\
lvelasco@ictp.it
\\[1mm]

\end{center}

\begin{abstract}
\noindent  We reconsider basic, in the sense of minimal field content, Pati-Salam $\times$ SU(3) family models which make use of the Type I see-saw mechanism to reproduce the observed mixing and mass spectrum in the neutrino sector. The goal of this is to achieve the observed baryon asymmetry through the thermal decay of the lightest right-handed neutrino and at the same time to be consistent with the expected experimental lepton flavour violation sensitivity. This kind of models have been previously considered but it was not possible to achieve a compatibility among all of the ingredients mentioned above. We describe then how different $SU(3)$ messengers, the heavy fields that decouple and produce the right form of the Yukawa couplings together with the scalars breaking the $SU(3)$ symmetry, can lead to different Yukawa couplings. This in turn implies different consequences for flavour violation couplings and conditions for realizing the right amount of baryon asymmetry through the decay of the lightest right-handed neutrino. Also a highlight of the present work is a new fit of the Yukawa textures traditionally embedded in $SU(3)$ family models.

\end{abstract}

\end{titlepage}

\newpage

\setcounter{footnote}{0}

\section{Introduction}

Some interesting models \cite{Dermisek:2006dc,Ross:2004qn} explaining fermion masses and mixing within the context of a GUT theory, a family symmetry and using Type I see-saw \cite{Minkowski:1977sc}, cannot easily account for the observed baryon asymmetry of the universe through thermal leptogenesis {\footnote{ It is possible to achieve the required baryon asymmetry through leptogenesis but only with the decay of the second lightest right-handed neutrino \cite{Vives:2005ra,Chun:2007ny}, not with the lightest \cite{Akhmedov:2003dg}.}.}

Working in the context of a gauge group Pati-Salam (PS) $ \times G_F$, where $G_F$ is a family group, we address two natural questions: how complicated is it to find a model realizing both mechanisms in a non-fine tuned way? and can we determine predictive features of such a model?

 Instead of concentrating on a particular model we first re-analyze, in the light of recent experimental measurements and analysis, the validity of the Yukawa textures that are commonly embedded in $SU(3)$ family models \cite{King:2003rf}-\cite{Calibbi:2009ja}. We find that while these Yukawa textures themselves are still compatible with experimental observations, the $SU(3)$ models that could potentially explain them should be modified. In particular, the messengers of such theories should be carefully analyzed since the details of them could determine the real predictability and viability of such models.

 This paper is organized as follows. In $\S$ 2 we state the parameters of the Yukawa textures that are usually embedded in $SU(3)$ family models \cite{King:2003rf}-\cite{Calibbi:2009ja}, briefly explaining which observables are used for the fit to experimental quantities. We leave for the Appendix A some details of the fits. One has been performed for  $y_b \ll y_t $, i.e. $\tan\beta$ small, (Table \ref{tbl:f_12_13_diffsgn_small_tbeta}) and the other one for $y_b \sim y_t $, i.e. $\tan\beta$ large, (Table \ref{tbl:f_12_13_diffsgn_large_tbeta}). The fit for large $\tan\beta$ is the preferred one.  Then in $\S$ 3 we re-state the assumptions for these textures to be embedded in $SU(3)$ family groups, paying attention to the messenger sector of the theory and giving as an example a minimal messenger content for the model in \cite{deMedeirosVarzielas:2005ax}. These messengers are not $SU(4)_{PS}$ singlets thus in principle this Ansatz could be embedded in $SO(10)$ models where the breaking to the PS group happens close to the unification scale, such that gauge coupling unification is not significantly altered.  In $\S$ 4 we explain how the observed neutrino mixing can arise in this particular example.  Under these guidelines we exemplify how different assumptions of the messenger masses can be taken into account to have different Yukawa couplings and different predictions for $\tan\beta$.

 The constraints coming from flavour violation and the parameters relevant for leptogenesis are discussed in $\S$ 5 and $\S$ 6 respectively. Lepton flavour violation gives important constraints on the elements of $Y^\nu$, the Yukawa coupling of the right-handed neutrinos in the basis where charged leptons are diagonal. As a result we find that for our example of messenger sector of an $SU(3)$ model, detailed in $\S$ 3, the messengers for the right-handed neutrinos and those for the rest of the fermions should have a rather different behaviour.

It is interesting to note that flavoured thermal leptogenesis
\cite{fl-lepto} can account for the explanation of the baryon
asymmetry in the universe.  The solution for low $\tan\beta$ is
not pretty sensitive to the element $Y^\nu_{11}$, while the
solution for high $\tan\beta$ it is. It was pointed out in
\cite{Antusch:2006cw} that in the context of right-handed neutrino
sequential dominance (RHNSD) \cite{King:RHND}, a ratio
$Y^\nu_{11}/Y^\nu_{21}\approx 0.1$ for a mass of the lightest
right-handed neutrino of $\mathcal{O}(10^{12})$ GeV and a relative
high $\tan\beta$ could account for the right value of the baryon
asymmetry through flavoured leptogenesis. Our solution for
explaining the mixing in the lepton sector differs from the RHNSD
solution and hence the weak dependence of $Y^\nu_{11}$ for low
$\tan\beta$.

  Finally we conclude in $\S$ 7 by summarizing that it is possible to construct a $SU(3)$ model with  Type I see-saw, account for the observed baryon asymmetry of the universe, with the decay of the lightest right-handed neutrino and satisfy flavour violation bounds. However, this requires a very complicated messenger structure that needs to be justified with further model building ingredients.


\section{Simple Yukawa Textures for fermions}

Yukawa textures of the form
\be
Y^f=
\left(
\begin{array}{ccc}
0 & a_{12} \ \epsilon^3_f & a_{13} \epsilon^3_f \\
 a_{21} \ \epsilon^3_f &  a_{22} \Yf \ \epsilon^2_f + a^\prime_{22} \  \epsilon^3_f & a_{23}\Yf \ \epsilon^2_f + a^\prime_{23} \  \epsilon^3_f \\
a_{31} \  \epsilon^3_f & a_{32} \ \epsilon^2_f + a^\prime_{32}\  \epsilon^3_f &  a_{33}
\end{array}
\right),\label{eq:yuk_o_text}
\ee
where the coefficients $a_{ij}={\mathcal{O}}(1)$,  $|a_{ij}|\approx |a_{ji}|$, $\epd={\mathcal{O}}(10^{-1})$ and $\epu \leq \epd$, have long been  considered to be a successful description of the mass eigenvalues and mixing in the quark sector \cite{Ross:2002fb}. There have been some experimental changes since the original fit of \cite{Roberts:2001zy} and the update in \cite{Chun:2007ny}; mainly the $\Delta M_{B_s}$ measurement \cite{Abulencia:2006ze}, already taken into account in \cite{Chun:2007ny}, and a better constraint of the angles of the unitary relation in the CKM matrix \cite{ckm_fit_09}.

In the Appendix A we present a result of the fit to  the observables;
\bea
&& V_{us}, \ V_{cb}, \ V_{ub}, \ \delta, \ \frac{m_u}{m_c},\ \frac{m_c}{m_t},\ \frac{m_d}{m_s},\ \frac{m_s}{m_b},
\label{exp:pr_to_fit}
\eea
where we use the standard notation of the CKM matrix \cite{Amsler:2008zzb} for the definition of $\delta=Arg[V_{ub}^*]$. It has been shown in Ref.~\cite{Roberts:2001zy} that expressions for these observables in terms of the Yukawa couplings of $u$ and $d$ sectors following the texture \eq{eq:yuk_o_text} have a rather simple form. We present some details and the results of the new fits for low and large $\tan\beta$ in Appendix A.

\section{Models based on $SU(3)$ family groups}

\subsection{Basic structure}
\begin{figure}[t]
\begin{center}
\begin{picture}(300,320)(0,0)
\ArrowLine(35,250)(85,250)
\DashArrowLine(85,290)(85,250){2}
\Line(83,292)(87,288)
\Line(87,292)(83,288)
\ArrowLine(110,250)(85,250)
\ArrowLine(110,250)(135,250)
\Line(108,252)(112,248)
\Line(112,252)(108,248)
\DashArrowLine(135,290)(135,250){3}
\Line(133,292)(137,288)
\Line(137,292)(133,288)
\ArrowLine(160,250)(135,250)
\ArrowLine(160,250)(185,250)
\Line(158,252)(162,248)
\Line(162,252)(158,248)
\DashArrowLine(185,290)(185,250){2}
\Line(183,292)(187,288)
\Line(187,292)(183,288)
\ArrowLine(235,250)(185,250)
\Text(65,242)[]{$\psi_i$}
\Text(85,300)[]{h}
\Text(100,242)[]{$\chi^{3i}_{R}$}
\Text(120,242)[]{$\bar\chi^{3}_{Ri}$}
\Text(135,301)[]{$\bar\phi^i_{3},$}
\Text(150,242)[]{$\chi^{1'}_{R}$}
\Text(170,242)[]{$\bar\chi^{1'}_{R}$}
\Text(185,301)[]{$\bar\phi^j_{3}$}
\Text(210,242)[]{$\psi_j^c$}
\ArrowLine(35,170)(85,170)
\DashArrowLine(85,210)(85,170){2}
\Line(83,212)(87,208)
\Line(87,212)(83,208)
\ArrowLine(110,170)(85,170)
\ArrowLine(110,170)(135,170)
\Line(108,172)(112,168)
\Line(112,172)(108,168)
\DashArrowLine(135,210)(135,170){3}
\Line(133,212)(137,208)
\Line(137,212)(133,208)
\ArrowLine(160,170)(135,170)
\ArrowLine(160,170)(185,170)
\Line(158,172)(162,168)
\Line(162,172)(158,168)
\DashArrowLine(185,210)(185,170){2}
\Line(183,212)(187,208)
\Line(187,212)(183,208)
\ArrowLine(235,170)(185,170)
\Text(65,162)[]{$\psi_i$}
\Text(85,220)[]{h}
\Text(100,162)[]{$\chi^{3i}_{R}$}
\Text(120,162)[]{$\bar\chi^{3}_{Ri}$}
\Text(135,221)[]{$\bar\phi^i_{123},$}
\Text(150,162)[]{$\chi^{1}_{R}$}
\Text(170,162)[]{$\bar\chi^{1}_{R}$}
\Text(185,221)[]{$\bar\phi^j_{23}$}
\Text(210,162)[]{$\psi_j^c$}
\ArrowLine(35,90)(85,90)
\DashArrowLine(85,130)(85,90){2}
\Line(83,132)(87,128)
\Line(87,132)(83,128)
\ArrowLine(110,90)(85,90)
\ArrowLine(110,90)(135,90)
\Line(108,92)(112,88)
\Line(112,92)(108,88)
\DashArrowLine(135,130)(135,90){3}
\Line(133,132)(137,128)
\Line(137,132)(133,128)
\ArrowLine(160,90)(135,90)
\ArrowLine(160,90)(185,90)
\Line(158,92)(162,88)
\Line(162,92)(158,88)
\DashArrowLine(185,130)(185,90){2}
\Line(183,132)(187,128)
\Line(187,132)(183,128)
\ArrowLine(235,90)(185,90)
\Text(65,82)[]{$\psi_i$}
\Text(85,140)[]{h}
\Text(100,82)[]{$\chi^{3i}_{R}$}
\Text(120,82)[]{$\bar\chi^{3}_{Ri}$}
\Text(135,141)[]{$\bar\phi^i_{23},$}
\Text(150,82)[]{$\chi^{1''}_{R}$}
\Text(170,82)[]{$\bar\chi^{1''}_{R}$}
\Text(185,141)[]{$\bar\phi^j_{123}$}
\Text(210,82)[]{$\psi_j^c$}
\ArrowLine(10,10)(60,10)
\DashArrowLine(60,50)(60,10){2}
\Line(62,52)(58,48)
\Line(58,52)(62,48)
\ArrowLine(85,10)(60,10)
\ArrowLine(85,10)(110,10)
\Line(83,12)(87,8)
\Line(87,12)(83,8)
\DashArrowLine(110,50)(110,10){3}
\Line(108,52)(112,48)
\Line(112,52)(108,48)
\ArrowLine(135,10)(110,10)
\ArrowLine(135,10)(160,10)
\Line(133,12)(137,8)
\Line(137,12)(133,8)
\DashArrowLine(160,50)(160,10){2}
\Line(158,52)(162,48)
\Line(162,52)(158,48)
\ArrowLine(160,10)(185,10)
\ArrowLine(185,10)(210,10)
\Line(183,12)(187,8)
\Line(187,12)(183,8)
\DashArrowLine(210,50)(210,10){2}
\Line(208,52)(212,48)
\Line(212,52)(208,48)
\ArrowLine(260,10)(210,10)
\Text(40,2)[]{$\psi_i$}
\Text(60,60)[]{h}
\Text(75,2)[]{$\chi^{3i}_{R}$}
\Text(95,2)[]{$\bar\chi^{3}_{Ri}$}
\Text(110,60)[]{$\bar\phi^i_{23},$}
\Text(125,2)[]{$\chi^{1''}_{R}$}
\Text(145,2)[]{$\bar\chi^{1''}_{R}$}
\Text(160,60)[]{$\hat H$}
\Text(175,2)[]{$\chi^1_{R}$}
\Text(195,2)[]{$\bar\chi^1_{R}$}
\Text(210,60)[]{$\bar\phi^j_{23}$}
\Text(235,2)[]{$\psi_j^c$}
\end{picture}\\
\caption{The Yukawa couplings for fermion masses and mixing arise once the right-handed messenger fields are integrated out. \label{fig:yuk_from_right_mess}}
\end{center}
\end{figure}
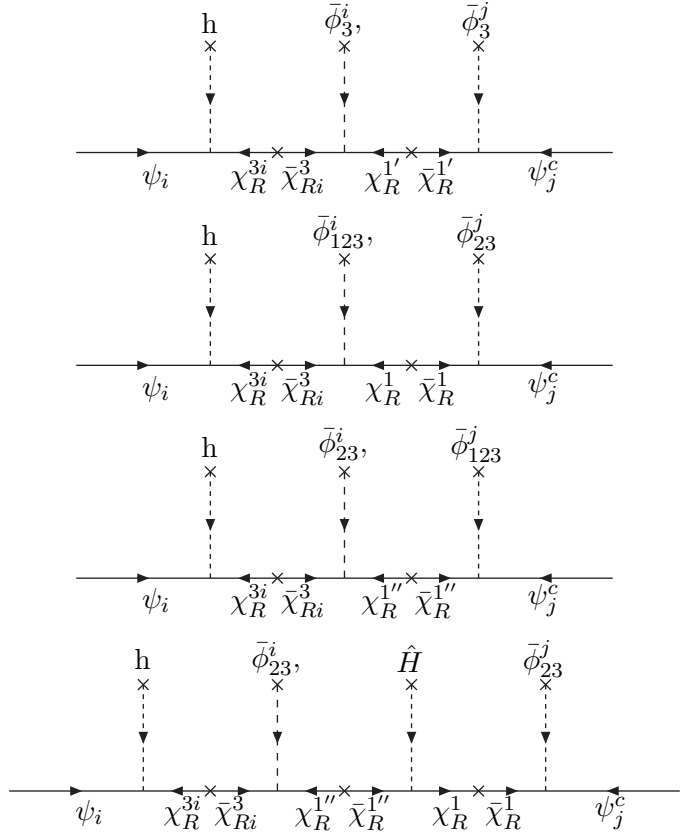

The form of the Yukawa couplings in \eq{eq:yuk_o_text} have been traditionally explained in the context of models with a $SU(3)$ family symmetry \cite{King:2003rf}-\cite{Calibbi:2009ja}. For definiteness we take the superpotential explaining the Yukawa couplings of quarks and charged leptons as in \cite{deMedeirosVarzielas:2005ax}:
\bea
W &=& \psi_i \frac{\overline\phi_3^i \overline\phi_3^j}{M_R^2}\psi^c_j H +
\psi_i \frac{\overline\phi_{23}^i H_{45} \overline\phi_{23}^j}{M_R^3}\psi^c_j H
+\psi_i \frac{\overline\phi_{123}^i \overline\phi_{23}^j }{M_R^2}\psi^c_j H+\psi_i \frac{\overline\phi_{23}^i \overline\phi_{123}^j}{M_R^2}\psi^c_j H,
\label{eq:W_SU3_e}
\eea
where the scalar components of the three superfields
$\overline\phi_3$,  $\overline\phi_{23}$, $\overline\phi_{123}$ (the so-called flavons or familions) break the $SU(3)$ symmetry by acquiring a vacuum expectation value (VEV) different from zero. Such VEVs are aligned in the flavour space as follows:
\bea
&&\langle \overline \phi_{3}\rangle =\left(0\quad 0\quad 1  \right) \times \left(\begin{array}{ccc}a_u & 0\\ 0 & a_d  \end{array}  \right),\quad
\langle \overline \phi_{23}\rangle=\left( 0\quad -1 \quad 1 \right) b\,,\nonumber\\
&& \langle \overline \phi_{123} \rangle=\left(1 \quad 1 \quad 1   \right) c\,.
\label{eq:vevs1}
\eea
Each of the fields above sits in an anti-triplet representation of $SU(3)$,
while the SM fermion fields, sit in triplet representations.
The field $\overline\phi_3$ is also a triplet of an $SU(2)_R$ group which could be
embedded in a GUT, while the others are PS or GUT singlets.
The superpotential of \eq{eq:W_SU3_e} also preserves two additional $U(1)$ symmetries
which have the role of forbidding dangerous contributions to specific Yukawa couplings.
In Ref.~\cite{deMedeirosVarzielas:2005ax} the specific choice of $U(1)$ charges can be
found.{\footnote{ We recall the reader that the vacuum alignment of flavons is crucial
in determining the order of magnitude of the vacuum expectation values of the flavons
appearing in \eq{eq:W_SU3_e}. But even this seems to be not enough to ensure a correct
description of fermion masses and mixing, that is why extra  $U(1)$'s need to be added.
It would be interesting to try to work out the correct size of Yukawa couplings just
through the vacuum alignment without the introduction of such extra $U(1)$'s.}}
In \eq{eq:W_SU3_e} $M_R$ indicates a common mass for the $\chi_R$ messengers.
Here we propose a specific messenger sector,
which the model of \cite{deMedeirosVarzielas:2005ax} does not specify,
that is compatible with the structure of \eq{eq:W_SU3_e} and
the form of the vacuum expectation values (VEVs) of the flavon fields in \eq{eq:vevs1}.
\begin{table}[t]
\begin{center}
\begin{tabular}{|c| c c c | c | c c|}
\hline
Field & $SU(4)_{\rm PS}$ & $SU(2)_{\rm L}$ & $SU(2)_{\rm R}$ & $SU(3)_{f}$ & $U(1)$ & $ U(1)^\prime$ \\
\hline
\hline
$\psi$ & $\bf{4}$ &  $\bf{2}$ &  $\bf{1}$ & $\bf{3}$ & $\bf{0}$ & $\bf{0}$ \\
$\psi^c$ & $\bf{\bar{4}}$ &  $\bf{1}$ &  $\bf{2}$ & $\bf{3}$ & $\bf{0}$ & $\bf{0}$ \\
\hline
$h$ & $\bf{1}$ &  $\bf{2}$ &  $\bf{2}$ & $\bf{1}$ & $\bf{-4}$ & $\bf{-4}$ \\
$\hat{H}$ & $\bf{15}$ &  $\bf{1}$ &  $\bf{3}$ & $\bf{1}$ & $\bf{2}$ & $\bf{2}$ \\
\hline
$\overline{\phi}_{123}$ & $\bf{1}$ &  $\bf{1}$ &  $\bf{1}$ & $\bf{\bar{3}}$ & $\bf{3}$ & $\bf{3}$ \\
$\overline{\phi}_{23}$ & $\bf{1}$ &  $\bf{1}$ &  $\bf{1}$ & $\bf{\bar{3}}$ & $\bf{1}$ & $\bf{1}$ \\
$\overline{\phi}_{3}$ & $\bf{1}$ &  $\bf{1}$ &  $\bf{3 \oplus 1}$ & $\bf{\bar{3}}$ & $\bf{2}$ & $\bf{2}$ \\
\hline
$\chi^1_R$ & $\bf{\bar{4}}$ &  $\bf{1}$ &  $\bf{2}$ & $\bf{1}$ & $\bf{1}$ & $\bf{1}$ \\
$\chi^{1\,\prime}_R$ & $\bf{\bar{4}}$ &  $\bf{1}$ &  $\bf{2}$ & $\bf{1}$ & $\bf{2}$ & $\bf{2}$ \\
$\chi^{1\,\prime\prime}_R$ & $\bf{\bar{4}}$ &  $\bf{1}$ &  $\bf{2}$ & $\bf{1}$ & $\bf{3}$ & $\bf{3}$ \\
$\chi^3_R$ & $\bf{\bar{4}}$ &  $\bf{1}$ &  $\bf{2}$ & $\bf{\bar{3}}$ & $\bf{4}$ & $\bf{4}$ \\
\hline
\hline
\end{tabular}
\end{center}
\caption{\label{tab:quantum-numbers} Charges of an example of messenger fields that could be used in the model of \cite{deMedeirosVarzielas:2005ax}.  }
\end{table}
Since our discussion concentrates on how the messenger sector could affect flavour changing processes and leptogenesis, we are going to discuss the details of it in the following sections. In Tab.~\ref{tab:quantum-numbers} we display the field-content of the model and the corresponding quantum numbers, without specifying additional fields needed by the breaking of the flavour symmetry and the vacuum alignment, which can be found in \cite{deMedeirosVarzielas:2005ax}. Since we assume that the Yukawa matrices arise mainly from the right-handed messenger sector sector, namely $M_{L}\gg M_{R}$, we have not included possible left-handed messengers (whose $U(1)$ charges would be the same as the right-handed messengers).
The field-content showed in Table \ref{tab:quantum-numbers} gives rise to the diagrams in Fig. \ref{fig:yuk_from_right_mess}.
Once the messenger fields are integrated out (by solving the supersymmetric equations of motion, $\partial_{X} W=0$,
$\partial_{\bar{X}} W=0$ \cite{Brizi:2009nn}), we get the following effective superpotential:
\be
W =  \psi_i \frac{\overline\phi_3^i \overline\phi_3^j}{M_{\chi^3_R} M_{\chi^{1 \prime }_R}}\psi^c_j h +
\psi_i \frac{\overline\phi_{23}^i \overline\phi_{23}^j}{M_{\chi^3_R} M_{\chi^{1}_R}}\psi^c_j h
\frac{\hat{H}}{M_{\chi^{1\prime \prime}_R}} 
+\psi_i \! \!\left[\frac{\overline\phi_{23}^i \overline\phi_{123}^j}{M_{\chi^3_R} M_{\chi^{1  }_R}}+\frac{\overline\phi_{123}^i \overline\phi_{23}^j}{M_{\chi^3_R} M_{\chi^{1 \prime \prime }_R}}\right] \! \! \psi^c_j h\,.
\label{eq:W_r-h_mess}
\ee
We did not write explicitly the $\mathcal{O}(1)$ couplings. In principle the messengers associated to the different kind of fermions, once the underlying GUT or PS groups are broken, could be different, hence from now on we denote the messenger masses as $M_R^f$, with $f=u,d,\nu$.  In particular it was pointed in \cite{Ross:2002fb} that $SU(2)_{\rm R}$ breaking effects are expected to split the messenger masses in the up and down sector, such that $Y^d$ and $Y^u$ can have different expansion parameters $\epd$ and $\epu$.

Let us now express the VEVs of the flavon fields and the messenger masses in terms of convenient expansion parameters, that we identify with those of \eq{eq:yuk_o_text} as follows:
\bea
&& \frac{\langle \overline \phi_{23}\rangle^2}{M^f_{\chi^3_R} M^f_{\chi^1_R}}=\epsilon^2_f\nonumber\\
&& \frac{\langle \overline \phi_{123} \overline \phi_{23} \rangle} {M^f_{\chi^3_R } M^f_{\chi^1_R}} =  \frac{\langle \overline \phi_{23} \overline \phi_{123} \rangle} {M^f_{\chi^3_R } M^f_{\chi^{1\prime \prime}_R}} =\epsilon_f^2\epsilon_d \nonumber\\
&& \frac{\langle \hat H \rangle}{M^f_{\chi^{1 \prime \prime }_R}}={\mathcal{Y}}^f.
\label{eq:vevs}
\eea
We have assumed that the VEV of the field $\hat H$ aligns in the direction of the hypercharge, ${\mathcal{Y}}^f$, of the given right-handed fermion.

After the breaking of $SU(3)$, the part of the effective K\"ahler potential corresponding to the matter fields is,
\bea
K_{\psi^{\dag}\psi} &=&  \psi^{\dag\, i} \psi_j \left[\delta_{i}^j  + \mathcal{O}\left(\frac{|\phi|^2}{M_P^2} \right) \right]\quad\mbox{and}
\label{eq:kahler1}\\
K_{\psi^{c\dag}\psi^c} &=&  \psi^{c\,\dag\, i} \psi^{c}_j \left[\delta_{i}^j + \frac{(\overline{\phi}_{123}^{\dagger})_i (\overline{\phi}_{123})^j }
{M_{\chi^{1\prime\prime}_R}^2} +
\frac{(\overline{\phi}_{23}^{\dagger})_i (\overline{\phi}_{23})^j } {M_{\chi^{1}_R}^2}
+\frac{(\overline{\phi}_{3}^{\dagger})_i (\overline{\phi}_{3})^j } {M_{\chi^{1\prime}_R}^2}\right. \nonumber \\
&&+\left. \mathcal{O}\left(\frac{|\phi|^2}{M_P^2}\right) \delta_{i}^j + ...
\right],
\label{eq:kahler2}
\eea
where the terms proportional to the hidden sector fields which break supersymmetry were not explicitly written. We have assumed that the left-handed messenger fields do not play a role in the determination of any physical quantity because the left-handed messengers are too heavy with respect to the right-handed ones:
\bea
r_M=\frac{M_{R^f}^2}{M_L^2}\ll 1.
\label{eq:dom_rh_mess}
\eea
\subsection{Canonical normalization}
As a consequence, the K\"ahler potential written above induces non-canonical kinetic terms only for the field $\psi^c$. In order to better identify these with physical quantities we redefine these superfields such that the kinetic terms are canonical:
\be
P^{\dag}_f K_f P_f = \bf{1}, \label{eq:can_transf_k}
\ee
with $f=\psi,\psi^c$.
Thus, the Yukawa matrices for canonically normalized fields are
\cite{King:2004tx, Antusch:2007re}:
\be
Y^f = P_f {\mathbf{Y}}^f P_{f^c},
\label{eq:can}
\ee
with $P_f\simeq {\bf 1}$, in our case. We use bold faces only for the Yukawa matrix in the non-canonical basis.
The choice of the matrices $P$ in \eq{eq:can_transf_k} is basis dependent, since if \eq{eq:can_transf_k} is also satisfied for $N^{\dag}_f K_f N_f = \bf{1}$ then $P$ and $N$ are related through a unitary matrix $U$: $P=N U$.  The effects of the canonical normalization does not depend on the choice of basis because one can show that the eigenvalues of the Yukawa matrices and the CKM mixing remain invariant independent of the choice of $P$ that satisfies the \eq{eq:can_transf_k}, but of course with the same assumptions on the messenger sectors \cite{King:2004tx,Antusch:2007re,Antusch:2008jf}. Furthermore as long as the K\"ahler metric  is of the form of \eq{eq:non_dg_kahler} and the Yukawa matrices as in \eq{eq:yuk_o_text} with all the coefficients $a_{ij}$ different from zero, the transformation of the matrix $P_{f^c}$ will not change the structure of the CKM mixing, it can just alter the $O(1)$ coefficients involved on it. However when some of the elements $a_{ij}$ are zero, the canonical normalization can have a rather different impact. We exemplify this with a choice of $P$ satisfying $P^{-1\dagger}=P^{-1}$.

From Eqs.~(\ref{eq:kahler2}, \ref{eq:yukf}), it is easy to check that such a transformation (acting only on right-handed superfields) induces a contribution to the the element $Y^f_{32}$ of the form:
\be
Y^f_{32} \sim {\mathbf{Y}}^f_{32} + {\mathcal{O}}({\mathbf{Y}}^f_{33}) \epsilon_f^2.
\label{eq:can23}
\ee
If we have $\mathbf{Y}^f_{33}\sim\mathcal{O}(1)$ then the second term of the expression above it is of the same order of the leading term in $\mathbf{Y}^f_{32}$, so that it does not spoil the pattern of fermion masses and mixing given by the Yukawa couplings with the structure of \eq{eq:yuk_o_text}. The most important changes in $Y^d$ and $Y^e$ are of the form:
\bea
Y^f\sim {\mathbf{Y}}^f +
\left(
\begin{array}{ccc}
{\mathcal{O}} (\epd^7) & {\mathcal{O}} (\epd^5) &- \frac{r_M^2}{2} {\mathcal{O}}({\mathbf{Y}}^f_{33})\epd^4  + {\mathcal{O}} (\epd^5)\\
{\mathcal{O}} (\epd^5) & -\frac{a_{23}}{2} \epd^4 & - \frac{r_M^2}{2} {\mathcal{O}}({\mathbf{Y}}^f_{33})\epd^2 +   {\mathcal{O}} (\epd^4)\\
- {\mathcal{O}}({\mathbf{Y}}^f_{33}) \frac{\epd^4}{2} & -\frac{{\mathcal{O}} ({\mathbf{Y}}^f_{33})}{2} \epd^2  & {\mathcal{O}} (\epd^4)
\end{array}
\right),
\eea
where we written $r_M$ just to emphasize that the induced changes in the upper and lower parts of the Yukawa matrix come from the contributions of the left- (very heavy) and right-handed messengers, respectively,  therefore are different. Then, the Yukawa matrices for $f=d,e$ (where $\epsilon_e=\epsilon_d$) in the canonical basis, to a good approximation, can expressed as
\be
Y^f\approx
\left(
\begin{array}{ccc}
-\frac{g_b^{\prime \prime}}{2}(g_{12}+g_{13}) \ \epsilon_d^7 & g_{12} \ \epsilon_d^3 & \frac{ g_{13} } {\rho^d } \ \epsilon_d^3 \\
 g_{21} \ \epsilon_d^3 &  g_{22} \Yf \ \epsilon^2_d + g^\prime_{22} \ \epsilon_d^3 & \frac{g_{23}}{\rho^d}\Yf \ \epsilon^2_d + \frac{g^\prime_{23}}{\rho^d} \ \epsilon_d^3 \\
(g_{31}-g_b^{\prime\prime}\frac{g^d_{33}}{2}\epd) \ \epsilon_d^3 & (g_{32}\Yf - g_b\frac{g^d_{33}}{2}) \ \epsilon^2_d + g^\prime_{32}\ \epsilon_d^3 &  \frac{g^d_{33}}{\rho^d}
\end{array}
\right),\label{eq:yukf}
\ee
where we have defined:
\bea
\rho^f=\sqrt{ 1+g_b^{\prime 2} \frac{a^2_f} {M^2_{\chi^{1\prime}_{R^f}}}}\,,
\eea
and the coefficients $g_{ij}$ are specified in \eq{eq:coeff_fud_op} in terms of the couplings of the fundamental operators in the superpotential of \eq{eq:renormW}, $g_b\ ,g_b^\prime$ and $g_b^{\prime\prime}$, which also appear in the
K\"ahler potential of \eq{eq:non_dg_kahler}{\footnote{Note that only terms in $a_f$ are
relevant in the K\"ahler potential in the diagonal components of
$K_{\psi^c \psi^{c\,\dag}}$ in \eq{eq:non_dg_kahler}, when taking the leading terms of
the canonical normalization for the matrices $P$ of \eq{eq:can_transf_k}.}}.
The expansion parameters $\epd$ and $\epu$ are specified in Appendix \ref{Ap:FIT} and are of
$\mathcal{O}(10^{-1})$ and  $\mathcal{O}(10^{-2})$ respectively.
We can see that the only effective changes of the canonical normalization are
in the elements $Y^f_{11}$, $Y^f_{23,32}$, $Y^f_{13}$ and $Y^f_{33}$ ($f=e,d$).
For the $Y^u$ matrix we have
\bea
Y^u\!=\!
\left(
\begin{array}{ccc}
-\frac{g_b^{\prime\prime}}{2}(g_{12}+g_{13})\ \epu^4 \epd^3 & g_{12} \ \epsilon_d \epsilon^2_u & \frac{g_{13}}{\rho^u} \ \epsilon_d\epsilon^2_u \\
 g_{21} \ \epsilon_d \epsilon^2_u &  g_{22} \Yf \ \epsilon^2_u + g^\prime_{22} \ \epsilon_d \epsilon^2_u & \frac{g_{23}}{\rho^u}\Yf \ \epsilon^2_u + \frac{g^\prime_{23}}{\rho}\ \epsilon_d \epsilon^2_u \\
(g_{31}\ -g_b^{\prime\prime}\frac{g^u_{33}}{2}\epd)\ \epsilon_d \epsilon^2_u  & (g_{32}\Yf - g_b\frac{g^u_{33}}{2}) \ \epsilon^2_u + g^\prime_{32}\ \epsilon_d \epsilon^2_u &  \frac{g^u_{33}}{\rho^u}
\end{array}
\right)\label{eq:yuk_u}
\eea
where again we have similar changes to those in the $d$ sector but  the only effective changes are in $Y^u_{23,32}$, $Y^u_{13}$ and $Y^u_{33}$ since $\epu< \epd$.

A real problem could arise in the neutrino sector. The leading term in the neutrino Yukawa coupling $Y^\nu$ is given by ${\mathbf{Y}}^\nu_{32}\sim \epsilon_d \epsilon_\nu^2$, since $\mathcal{Y}(\nu_R)=0$, which it is required by the large mixing in the Pontecorvo-Maki-Nakagawa-Sakita (PMNS) matrix. However, the transformation of \eq{eq:can} to the canonical basis induces a dominant contribution to  $Y^\nu_{32}$:
\bea
Y^\nu=
\left(
\begin{array}{ccc}
-\frac{g_b^\prime}{2}(g_{12}+g_{13})\ \epn^4 \epd^3 &  g_{12} \ \epn^2 \epd &  \frac{g_{13}}{\rho^\nu} \ \epn^2 \epd \\
g_{21} \ \epn^2 \epd & g^\prime_{22} \ \epn^2 \epd   & \frac{g^\prime_{23}}{\rho^\nu} \ \epn^2 \epd\\
g_{31} \ \epn^2 \epd & g^\prime_{32} \ \epn^2 \epd -g_b\frac{g^\nu_{33}}{2}\epn^2 & \frac{g^\nu_{33}}{\rho^\nu}
\end{array}
\right),\label{eq:yuk_nu}
\eea
which spoils the structure of the light neutrino mass matrix for $g^\nu_{33} \sim\mathcal{O}(1)$ if we would like to use the solutions for the PMNS mixing as in \cite{deMedeirosVarzielas:2005ax}.
However, as long as we put the constraint of
\begin{equation}
g_b g^\nu_{33} < \epsilon_d,
\label{eq:ynu3}
\end{equation}
the effect of the canonical normalization for the proposed messenger sector leads to the right structure of $Y^\nu$ with a solution that we detail in the following section.

Note however that \eq{eq:ynu3} it is not a relation between the canonical Yukawa couplings which can lead more directly to the mass eigenvalues. From \eq{eq:W_r-h_mess} we identify the couplings $g^f_{33}$ in the non-canonical basis: $g^f_{33}=g_a' g_b'\frac{\langle \overline\phi_3^2 \rangle}{M_{\chi^3_R} M_{\chi^{1 \prime }_R}}= g_a' g_b' \frac{a_f^2}{M_{\chi^3_R} M_{\chi^{1 \prime }_R}}$ and then in the canonical basis we have
\bea
Y^f_{33}=  g_a^{\prime} g_b^{\prime}
\frac{\frac{a^2_f} {M^f_{\chi^3_R} M_{\chi^{1\prime}_{R^f}} }}{\sqrt{1+ g^{\prime 2}_b \frac{a^2_f}{M^2_{\chi^{1\prime}_{R^f}}} }}=
\left\{
\begin{array}{l}
\frac{g_a^{\prime} g_b^{\prime }}{\sqrt{1+g_b^{\prime 2}}} \frac{a_f}{M^f_{\chi^3_R}}\;\; \text{for} \ f=u, d,e \ (a_e=a_d) \\
g_a^{\prime} g_b^{\prime} \frac{a^2_u}{M^f_{\chi^3_R} M_{\chi^{1\prime}_{R^\nu}}}\;\; \text{for} \ f=\nu
\end{array}
\right.,
\eea
assuming that $a^f$ are real {\footnote{This effect on the eigenvalues after canonical normalization is independent on the choice of $P$. This is easily checked by noting that $\rm{Det}[Y^f]=\rm{Det}[\mathbf{Y}^f]\rm{Det}[P_{f^c}]$ is independent of the choice of $P_{f^c}$. Then given the structure of \eq{eq:non_dg_kahler} we have $\rm{Det}[P]\approx \left[(1+g^{''2}_b \epd^2 \epsilon_f^2)^{1/2}(1+g^{2}_b \epsilon_f^2)^{1/2}(1+g^{\prime 2}_b \frac{a^2_f}{M^2_{\chi^{1\prime}_{R^f}}})^{1/2}  \right]^{-1}$}}. Only the first expression of $Y^f_{33}$ after the equal sign is general, the second expression assumes
\bea
M_{\chi^{\prime}_{R^f}}=a_f,\quad f=u,d,e;\quad  M_{\chi^{\prime}_{R^\nu}} \gg a_u.
\label{eq:assump_messneuts}
\eea
The last inequality is compatible with \eq{eq:ynu3} since we need
\bea
M_{\chi^{1\prime}_{R^\nu}}> g_a'g_b'g_b\frac{a_u}{\epd},
\label{eq:newconst}
\eea
and we have assumed the first equality in \eq{eq:assump_messneuts}. Note that in order to satisfy \eq{eq:vevs} we need also to assume that $ M_{\chi^{1\prime}}=M_{\chi^{1\prime\prime}} $. In this case then the Yukawa matrix $Y^\nu$ is effectively insensitive to the canonical normalization. The canonical couplings of course will correspond mainly to the physical couplings of the third generation because the diagonalization of the matrices will not affect them too much.
 Then we can work out some interesting relations by connecting the scale $M^u_{\chi^3_R}$
 with the different sectors and using the relations (\ref{eq:vevs}):
\bea
 \frac{\epn^2}{\epu^2}=\frac{M_{\chi^{1}_{R^u}}}{M_{\chi^{1}_{R^\nu}}},&&
\frac{\epu^2}{\epd^2}=\frac{M_{\chi^{1}_{R^d}}}{M_{\chi^{1}_{R^u}}}.
\label{eq:relofphys_aftcn}
\eea
On the other hand for a physical top Yukawa coupling we need $y_t\approx \mathcal{O}(1) \approx Y^u_{33}$ and hence we conclude also that we need $M^u_{\chi^{3}_{R}}\approx a_u$ and since the couplings in \eq{eq:newconst} are of ${\mathcal{O}}(1)$, the requirement on the mass of ${M_{\chi^{1\prime}_{R^\nu}}}$ can be understood simply in terms of $a_u$ and $\epd$, and there is enough freedom to choose it at the right order without affecting the other scales which, as we have seen, should be similar to each other. The second relation of \eq{eq:vevs} can be also used to get some relations among masses but it is not useful since it introduces other unknown parameter: $c$.
Note from \eq{eq:relofphys_aftcn} that, while there is a relation between ${M_{\chi^{1}_{R^u}}}$ and ${M_{\chi^{1}_{R^d}}}$ through parameters that can be determined from Yukawa couplings, there is not a such a relation between $a_u$ and $a_d$ and hence there is some freedom to set the value of $\tan\beta$:
\bea
\tan\beta=\frac{v_u}{v_d}
= \frac{m_t(M_Z)}{m_b(M_Z)}  \frac{a_d}{a_u} ~
\frac{M^u_{\chi^3_{R}}}{M^d_{\chi^3_{R}}}.
\label{eq:tanbeta_det}
\eea
From the Eqs.~(\ref{eq:kahler1},\ref{eq:kahler2}) for the K\"ahler potential,
we can easily derive the structure of the soft-mass matrices as they appear
at the SUSY-breaking scale and after the canonical normalization:
\bea
m^2_{\tilde{f} =\tilde Q, \tilde L}&\simeq&
\left(
\begin{array}{ccc}
1 & & \\
 & 1& \\
 & & 1\\
\end{array}
\right) m^2_0 +
r_M
\left(
\begin{array}{ccc}
 \epsilon^{\prime\,2}_f& \epsilon^{\prime\,2}_f & \epsilon^{\prime\,2}_f \\
 \epsilon^{\prime\,2}_f & \epsilon^{2}_f & \epsilon^{2}_f \\
\epsilon^{\prime\,2}_f  & \epsilon^{2}_f & {\mathcal{Y}}^f_{33}
\end{array}
\right) m^2_0, \\
m^2_{\tilde f^c}&\simeq&
\left(
\begin{array}{ccc}
1 & & \\
 & 1& \\
 & & 1\\
\end{array}
\right) m^2_0 +
\left(
\begin{array}{ccc}
 \epsilon^{\prime\,2}_f& \epsilon^{\prime\,2}_f & \epsilon^{\prime\,2}_f \\
 \epsilon^{\prime\,2}_f & \epsilon^{2}_f & \epsilon^{2}_f \\
\epsilon^{\prime\,2}_f  & \epsilon^{2}_f & {\mathcal{Y}}^f_{33}
\end{array}
\right) m^2_0 \,.\label{eq:soft}
\eea
Here $\epsilon^\prime_f \simeq \epsilon_f \epsilon_d$. The expressions for $m^2_{\tilde Q, \tilde L}$ clearly just depend only on the possible left-handed messengers, we are writing on the right-hand side of \eq{eq:soft} expressions in terms of the $\epsilon_f$ parameters of the Yukawa matrices and that are dominated by the right-handed messengers, hence we could equally put $\epsilon_u$ or $\epsilon_d$ multiplied by its corresponding right-handed mass.

\subsection{Effective flavour matrices}
After the rotation to the so-called super CKM (SCKM) basis, where the corresponding Yukawa matrix is diagonal, \eq{def:diagz_matxs}, the soft squared mass matrices take the form:
\bea
m^2_{\tilde f}&\simeq&
\left(
\begin{array}{ccc}
1 & & \\
 & 1& \\
 & & 1\\
\end{array}
\right) m^2_0 +
r_M
\left(
\begin{array}{ccc}
 \epsilon_d \epsilon^{2}_f& \epsilon_d \epsilon^{2}_f & \epsilon_d \epsilon^{2}_f \\
 \epsilon_d \epsilon^{2}_f & \epsilon^{2}_f & \epsilon^{2}_f \\
\epsilon_d \epsilon^{2}_f  & \epsilon^{2}_f & {\mathcal{Y}}^f_{33}
\end{array}
\right) m^2_0, \label{eq:softSCKM-LL}\\
m^2_{\tilde f^c}&\simeq&
\left(
\begin{array}{ccc}
1 & & \\
 & 1& \\
 & & 1\\
\end{array}
\right) m^2_0 +
\left(
\begin{array}{ccc}
\epsilon_d \epsilon^{2}_f & \epsilon_d \epsilon^{2}_f & \epsilon_d \epsilon^{2}_f \\
\epsilon_d \epsilon^{2}_f & \epsilon^{2}_f & \epsilon^{2}_f \\
\epsilon_d \epsilon^{2}_f  & \epsilon^{2}_f & {\mathcal{Y}}^f_{33}
\end{array}
\right) m^2_0 \,.\label{eq:softSCKM-RR}
\eea
These are the relevant quantities that we use to compute the flavour violating parameters from which we can immediately see that flavour violating off-diagonal parameters are present even before $M_{\text{GUT}}$ \cite{King:2003rf,Ross:2004qn}, at the scale at which the flavour symmetry is broken.

Let us finally comment that in the limit $M_{L}\gg M_{R}$ that we are considering (as well as in absence of left-handed messengers), there is no flavour violation in the left-handed sfermion sector above $M_{\text{GUT}}$, only in the right-handed sector. Still of course, flavour violating terms in $m^2_{\tilde f}$ can be generated by the RGE running down to the EW scale, as we will see in $\S$ \ref{sec:lfv}.

From the discussion in this section we have learned that the Yukawa couplings obtained from $SU(3)$ models with the chosen messenger sector are compatible with the following form of Yukawa matrix
\be
Y^f=
\left(
\begin{array}{ccc}
0 & a_{12} \ \epsilon_d \epsilon^2_f & a_{13} \epsilon_d \epsilon^2_f \\
 a_{21} \ \epsilon_d \epsilon^2_f &  a_{22} \Yf \ \epsilon^2_f + a^\prime_{22} \ \epsilon_d \epsilon^2_f & a_{23}\Yf \ \epsilon^2_f + a^\prime_{23} \ \epsilon_d \epsilon^2_f \\
a_{31} \ \epsilon_d \epsilon^2_f & a_{32}\Yf \ \epsilon^2_f + a^\prime_{32}\ \epsilon_d \epsilon^2_f &  a^f_{33}
\end{array}
\right),\label{eq:yuk_o_as_int_insu3}
\ee
where ${\mathcal{Y}}^\nu=0$, $a^\nu_{33}< \epd $, and all other coefficients are ${\mathcal{O}}(1)$. We do not necessarily need to have $a_{ij}=\pm a_{ji}$ because after canonical normalization the relations between these ${\mathcal{O}}(1)$ can be altered. In order to identify this matrix with the matrix of \eq{eq:yuk_o_text} for $Y^u$ we have to make the redefinitions of coefficients from the fit in the Appendix A as follows
\bea
a^u_{1j}\rightarrow a^u_{1j}\frac{\epd}{\epu},
\eea
where in the left hand side are the coefficients of the matrix (\ref{eq:yuk_o_as_int_insu3}) and the right-hand side the coefficients of Table \ref{tbl:f_12_13_diffsgn_small_tbeta}. The coefficients $a^\prime_{22,23,32}$ cannot be really determined with this kind of fits. All of the other coefficients are as in Appendix A.


%
\section{Constraints from lepton mixing \label{sec:leptmix}}
The Yukawa matrices in the lepton sector can be diagonalized by the transformations: $Y^e=U^{e\dagger}_L \hat Y^e U^e_R$, $Y^\nu=U^{\nu\dagger}_L \hat Y^\nu U^\nu_R$, where $\hat Y^f$ are diagonal matrices. Since constraints from leptogenesis
are better understood in the basis where the charged leptons are diagonal we consider the Type I see-saw formula in this basis:
\bea
 \hat m^\nu_{LL} &=&
U^\nu_L \left(U^{e\dagger}_L Y^\nu \right) M^{-1}_R \left(Y^{\nu T} U^{e*}_L\right) U^{\nu T}_L\nonumber\\
&=& U^\nu_L \widetilde{Y}^{\nu} M^{-1}_R \widetilde{Y}^{\nu T} U^{\nu T}_L=
U_{\text{PMNS}}^\dagger  m^\nu_{LL} U_{\text{PMNS}}^*,
\eea
where  $\hat m^\nu_{LL}$ is the diagonal matrix of neutrino mass eigenvalues:
\bea
\hat m^\nu_{LL}=m_3\left(
\begin{array}{ccc}
\frac{m_1}{m_3} &      0          & 0\\
0               & \frac{m_2}{m_3} & 0\\
0               &                 & 1
\end{array}
 \right).
\eea
We can then define
\bea
q_1 \equiv \frac{m_1}{m_3},\quad  q_2 \equiv \frac{m_2}{m_3}
\eea
and use the standard form of the PMNS matrix \cite{Amsler:2008zzb} to determine the form of $m^\nu_{LL}$  in terms of $q_i$ and the mixing angles $\theta_1$, $\theta_2$ and $\theta_3$:
\bea
m^\nu_{LL}=m_3 U_{\text{PMNS}}
\left(
\begin{array}{ccc}
q_1 &      0          & 0\\
0               & q_2 & 0\\
0               &                 & 1
\end{array}
 \right)
U_{\text{PMNS}}^{T},
\eea
which can be used to quickly compare it to the form of $m^\nu_{LL}$ given
by a particular model. In the following we will be referring just to $\widetilde{Y}^{\nu}$,
taking the form of $Y^{\nu}$ from \eq{eq:yuk_o_as_int_insu3}, then we can express it as
\bea
\widetilde{Y}^\nu=
\left(
\begin{array}{ccc}
-\frac{a^e_{12}}{y_\tau} a^\nu_{21} \epd^2\epn^2 & a^\nu_{12} \epd\epn^2 & a^\nu_{13} \epd\epn^2\\
a^\nu_{21} \epd\epn^2 & a^\nu_{22} \epd\epn^2  & a^\nu_{23} \epd\epn^2 \\
a^\nu_{31} \epd\epn^2 & a^\nu_{32}  \epd\epn^2  & Y^\nu_{33} \\
\end{array}
\right)
+
\left(
\begin{array}{ccc}
0 & b_{12} \epsilon^2_\nu\epsilon_d^2 & b_{13} \epsilon^2_\nu\epsilon_d^2 \\
b_{21} \epsilon^2_\nu\epsilon_d^2 & b_{22} \epsilon^2_\nu\epsilon_d^2 & b_{23} \epsilon^2_\nu\epsilon_d^2\\
b_{31} \epsilon^2_\nu\epsilon_d^2 & b_{32}\epsilon^2_\nu\epsilon_d^2 &  b_{33} \epsilon^2_\nu\epsilon_d^2
\end{array}
\right)
\label{eq:ynu_eff}
\eea
which has acquired an effective $Y^\nu_{11}$ element different from zero but leaving
the structure unchanged in the other elements at leading order,
as it can be seen from \eq{eq:yuk_o_as_int_insu3}.
From now on we call
\bea
a^\nu_{11}=-a^e_{12} a^\nu_{21}/y_\tau.
\label{eq:rel_anu11}
\eea
The sub-dominant contribution, the second term of the right-hand side of \eq{eq:ynu_eff}, comes entirely
from the change of basis where the charged leptons are diagonal.
For the hierarchy $q_1 < q_2<1$ we can better see the structure of the matrix
$m^\nu_{LL}$ in the limit $s_{12}=1/{\sqrt{3}}$, $s_{23}=1/{\sqrt{2}}$:
\bea
m^\nu_{LL}&=&m_3
\left(
\begin{array}{lll}
 \frac{2 q_1}{3}+\frac{q_2}{3} & \frac{q_2}{3}-\frac{q_1}{3} &
   \frac{q_1}{3}-\frac{q_2}{3} \\
 \frac{q_2}{3}-\frac{q_1}{3} &
   \frac{q_1}{6}+\frac{q_2}{3}+\frac{1}{2} &
   -\frac{q_1}{6}-\frac{q_2}{3}+\frac{1}{2} \\
 \frac{q_1}{3}-\frac{q_2}{3} &
   -\frac{q_1}{6}-\frac{q_2}{3}+\frac{1}{2} &
   \frac{q_1}{6}+\frac{q_2}{3}+\frac{1}{2}
\end{array}
\right)\nonumber\\
&+& m_3
\left(
\begin{array}{lll}
s_{13}^2 & \frac{s_{13}}{\sqrt{2}}  -\frac{ q_2 s_{13}}{3\sqrt{2}} &
\frac{s_{13}}{\sqrt{2}} -\frac{ q_2 s_{13}}{3\sqrt{2}}\\
\frac{s_{13}}{\sqrt{2}} -\frac{ q_2 s_{13}}{3\sqrt{2}} &- \frac{\sqrt{2}}{3} q_2 s_{13} & \frac{1}{6} q_2 s_{13}^2\\
\frac{s_{13}}{\sqrt{2}}  -\frac{ q_2 s_{13}}{3\sqrt{2}} &
\frac{1}{6} q_2 s_{13}^2 &  \frac{\sqrt{2}}{3} q_2 s_{13}
\end{array}
\right).
\label{eq:maj_req}
\eea
We can see that the dominant contribution for the determination
of the $\theta_{23}$ and $\theta_{12}$ mixing angles comes from the first term in \eq{eq:maj_req}.
The second term determines the size of the $\theta_{13}$ angle.
Comparing \eq{eq:maj_req} to the see-saw formula $m^{\nu}_{LL} =  -\frac{v_u^2}{2} \tilde Y^\nu M^{-1}_R{\tilde Y^{\nu T}}$, we see that terms proportional to
$1/M_1$ must dominate and we need the following hierarchy of right-handed neutrino masses:
\bea
\frac{\epn^4\epd^2}{(Y^\nu_{33})^2} M_3 >  M_2> M_1.
\label{eq:cond_1_M1_M2}
\eea
Remember that from constraints on canonical normalization we require  $Y^\nu_{33}\leq  \epd $ (see \eq{eq:yuk_nu} and comments below).
Let us consider first term of \eq{eq:maj_req}, ignoring for the moment the second term and
assuming that the leading contributions in $1/M_1$ in the $(2,3)$ sub-matrix of $m^\nu_{LL}$
generate the atmospheric mixing. The solar mixing angle and $q_2$ are explained through the terms which go like $1/M_2$ in $m^\nu_{LL}$.

 With the form of $\tilde Y^{\nu}$ of \eq{eq:ynu_eff}, and considering at first approximation $M_R$ to be diagonal we can compare the see-saw formula $m^{\nu}_{LL} =  -\frac{v_u^2}{2} \tilde Y^\nu M^{-1}_R{\tilde Y^{\nu T}}$ to the mass matrix in \eq{eq:maj_req} in order to put conditions on the elements on $\tilde Y^\nu$ and $M_R$ to achieve the desired mixing in the lepton sector, then we have:
\bea
\frac{M_1}{2 a^{\nu 2}_{31} \epd^2\epn^4}%
\left(
\begin{array}{c}
\frac{a^\nu_{12} a^\nu_{22} \epd^2\epn^4}{M_2}+\frac{a^\nu_{11} a^\nu_{21} \epd^3\epn^4}{M_1}\\
\frac{a^\nu_{12} a^\nu_{32} \epd^2\epn^4}{M_2}+\frac{a^\nu_{11} a^\nu_{31} \epd^3\epn^4}{M_1}\\
\frac{a^2_{22} \epd^2\epn^4}{M_2}
\end{array}
\!\!\right)=
\left(
\begin{array}{c}
1\\
-1\\
1\\
\end{array}
\!\!\right)\frac{q_2}{3},\quad \
\frac{M_1}{2 a^{\nu 2}_{13}  \epd^2\epn^4}
\left(\!\!
\begin{array}{c}
\frac{a^2_{21} \epd^2\epn^4}{M_1}\\
\frac{a^\nu_{21} a^\nu_{31} \epd^2\epn^4}{M_1}\\
\frac{a^{\nu 2}_{31}  \epd^2\epn^4}{M_1}
\end{array}
\!\!\right)=
\left(
\begin{array}{c}
1\\
1\\
1\\
\end{array}
\right)\frac{1}{2},\nonumber
\eea
which implies
\bea
\label{eq:C1-cn2_maj}
&& a^{\nu 2}_{21}=a^\nu_{21}a^\nu_{31}=a^{\nu 2}_{31},\nonumber\\
&& a^\nu_{22}= -a^\nu_{23} \,.
\eea
On the other hand we also need
\bea
M_1>M_2 \epd \frac{a^\nu_{11}a^\nu_{21}}{a^\nu_{22}a^\nu_{12}},
\label{eq:cond_2_M1_M2}
\eea
in order to be compatible with \eq{eq:C1-cn2_maj}, but there is only a narrow range of $M_2$ and $M_1$ that can simultaneously satisfy \eq{eq:cond_1_M1_M2}.
In this case we have
\bea
m_3 &\sim & \epn^4 \epd^2  \frac{v^2_u}{M_1}, \nonumber\\
m_2 &\sim & \epn^4 \epd^2  \frac{v_u^2}{M_2}.
\label{eq:m3nu}
\eea
Then from $m_3$ in \eq{eq:m3nu}, the order of magnitude of $M_1$ then can be fixed as follows:
\bea
M_1\approx 1.7 \sin^2\beta \left[\frac{\epd}{0.13}\right]^2 \left[\frac{\epn}{0.3}\right]^4 \left[\frac{0.05~\text{eV}}{m_3} \right] \left[\frac{v}{246~\text{GeV}} \right]^2 \times 10^{11} ~\text{GeV}.
\label{eq:M1_nr}
\eea
Since there is just an upper experimental limit on the value of $\theta_{13}$ and no constraints on $q_1$, the value of $M_3$ is not really further restricted other than from \eq{eq:cond_1_M1_M2}. Then if it is sufficiently large, the Yukawa coupling $Y^\nu_{33}$ does not play a significant role into the determination of the parameters in \eq{eq:m3nu}.  However it is nice to see that from the requirement of $Y^\nu_{33}\leq \epd$ we can go ahead and propose that
\bea
&& a^\nu_{33}=\epd \epn^2 b_{33},\nonumber\\
&& M_3\sim \frac{M_1}{\epd} ,\nonumber\\
&& a^\nu_{23} = -b_{33}.
\label{eq:C1-cn3_maj}
\eea
With the assumptions above (\ref{eq:C1-cn3_maj}) we then have
\bea
m_1 &\sim & \epn^4 \epd^2  \frac{v_u^2}{M_3}.
\label{eq:m1nu}
\eea
To summarize, $q_1$ is now described by the sub-dominant terms in $1/M_1$ in the elements $(1,2)-(2,1)$, $(1,3)-(3,1)$ and the terms in $1/M_3$ in all elements of  $m^\nu_{LL}$. This also assumes \eq{eq:m1nu} and of course we have taken into account the second term of \eq{eq:maj_req} which gives small variations to the requirements of the coefficients in \eq{eq:C1-cn2_maj} and \eq{eq:C1-cn3_maj}. The mixing angle $\theta_{13}$ is restricted by this choice and by the coefficient $a^\nu_{11}$. The values of the coefficients of $Y^\nu$ and the outputs of the mass eigenvalues and mixing angles are given in Table \ref{tbl:neut_par}.

\section{Constraints from LFV decays}\label{sec:lfv}
\begin{table}[t]
\begin{center}
\begin{tabular}{|c| c c c |}
\hline\hline
\multicolumn{4}{|c|}{Example of flavour violating parameters for $M_R\sim {\mathcal{O}}(M_{\Gt}) $}\\
\hline\hline
$(\delta^e_{\rm XY})_{ij}$ & LL & RR & LR  \\
\hline
\hline
$ij=12$ & $r_M\frac{\epsilon_d^3}{\mathcal{R}} \sim 0$ & $\frac{\epsilon_d^3}{\mathcal{R}} \lesssim 10^{-3}$ &  $\frac{v}{\sqrt{1+\tan\beta^2}}
~\frac{A_0 \epsilon_d^3}{{\mathcal{R}} m_0^2} \lesssim 10^{-5}$           \\
$ij=13$ & $r_M\frac{\epsilon_d^3}{\mathcal{R}} \sim 0$ & $\frac{\epsilon_d^3}{\mathcal{R}} \lesssim 10^{-3} $ &  $\frac{v}{\sqrt{1+\tan\beta^2}}
~ \frac{A_0 \epsilon_d^3}{{\mathcal{R}} m_0^2}\lesssim 10^{-5}$  \\
$ij=23$ & $r_M\frac{\epsilon_d^2}{\mathcal{R}}  \sim 0$ & $\frac{\epsilon_d^2}{\mathcal{R}} \lesssim 10^{-2} $ &  $\frac{v}{\sqrt{1+\tan\beta^2}}
 ~\frac{A_0 \epsilon_d^2}{\mathcal{R} m_0^2}\lesssim 10^{-4}$ \\
\hline
\hline
\end{tabular}
\end{center}
\caption{\label{tab:flav_violt_pars} Leptonic flavour violating parameters.
The running to the EW scale is parametrized by the running factor ${\mathcal{R}}$, which depends on the relations at the GUT scale between $M_{1/2}$ and $m_0$. While flavour parameters of the type $(\delta^f_{\rm XX})_{ij}$ do not depend heavily on the details of the supersymmetric parameters, the ones like $(\delta^f_{\rm LR})_{ij}$ do.  The numbers quoted for this later parameters correspond to the supersymmetric point $\left\{A_0,m_0\right\}=\left\{-520,370\right\}$ GeV, $\tan\beta= 50$, following \cite{Olive:2008vv}. We present this example to contrast it with the flavour violating parameters generated by our example of $\S$ 3 .}
\end{table}
We have seen in $\S$ 3 that, in the basis where charged leptons are diagonal, the soft mass matrices of
Eqs.~(\ref{eq:softSCKM-LL}, \ref{eq:softSCKM-RR}) have
flavour violating off-diagonal entries already at the flavour symmetry breaking scale.
Moreover, the trilinear $A^f$ matrices are in general not aligned with the corresponding Yukawa matrices
\cite{Ross:2004qn,Antusch:2007re,Calibbi:2008qt}. As a consequence, sources of flavor mixing are expected to arise from all sectors:
LL, RR and LR.

From Eqs.~(\ref{eq:softSCKM-LL}, \ref{eq:softSCKM-RR}), we see that the leptonic flavour violating mass insertions (MI) for the
$\mu-e$ transitions are:
\be
(\delta^e_{LL})_{12}~ \simeq ~  r_M \,\frac{ \epsilon_d^3}{\mathcal{R}}\,,
~~(\delta^e_{RR})_{12}~ \simeq ~ \frac{\epsilon_d^3}{\mathcal{R}}\,,
\ee
where the factor $\mathcal{R}$ accounts for the running of the slepton masses. If the right-handed neutrinos had decoupled above or at the GUT scale, the present experimental bound, ${\rm BR}(\mu\to e\gamma)< 1.2 \times 10^{-11}$ \cite{exp:muegamma}, could have been easily satisfied. Since $\mathcal{R}$ from the GUT to the EW scale is simply given by
\be
\mathcal{R}\simeq
\left\{\begin{array}{c}
\frac{m_0^2 + 0.5 M_{1/2}^2}{m_0^2} ~~~ \mathrm{LH~sleptons} \\
\frac{m_0^2 + 0.15 M_{1/2}^2}{m_0^2} ~~~ \mathrm{RH~sleptons}
\end{array}\right.,
\ee
and hence in the limit we are taking, $M_L \gg M_R$ ($r_M\rightarrow 0$), we have
\be
(\delta^e_{LL})_{12}~ \simeq ~ 0\,,~~(\delta^e_{RR})_{12}~ \simeq ~ \mathcal{O}(10^{-3})\,.
\ee
 Indeed, this just would give a weak constraint to the SUSY parameter space, excluding regions with light slepton and gaugino masses \cite{Masina:2002mv}. A larger source of lepton flavour violation (LFV) is in any case provided by the LR mass insertions, in the case of non-vanishing $A_0$. In Table \ref{tab:flav_violt_pars}, we present an estimate of all the leptonic MIs, as predicted by the $SU(3)$
flavour symmetry and by the choices made about the messenger fields for an example as if the right-handed neutrinos had decoupled above the GUT scale.

Now we take into account also the running of the off-diagonal elements of $m^2_{\tilde L}$,  driven by the neutrino Yukawa couplings \cite{borzumati-masiero}. This well known effect can be estimated to be in leading-log approximation \cite{Hisano:1995cp,Hisano:1995nq} as:
\be
(m^2_{\tilde L})_{i\neq j}(M_Z) \simeq (m^2_{\tilde L})_{i\neq j}(M_{\rm GUT}) - \frac{3 m_0^2 + A_0^2}{8 \pi^2} \left(Y^\dag_\nu
\ln \frac{M_{\rm GUT}}{M_R} Y_\nu\right)_{i\neq j}\,.
\ee
As a consequence, considering only the contribution from the running, we get
\bea
(\delta^e_{LL})_{12} & \!\simeq &  - \frac{1}{8 \pi^2} \epsilon_\nu^4 \epsilon_d^2
\left[\epsilon_d \ln \frac{M_{\rm GUT}}{M_{R_1}} + \ln \frac{M_{\rm GUT}}{M_{R_2}} + \ln \frac{M_{\rm GUT}}{M_{R_3}}\right] \frac{3 m_0^2 + A_0^2}{m_0^2 + 0.5 M_{1/2}^2}\,, \label{eq:delta12}\\
(\delta^e_{LL})_{23} & \!\simeq & - \frac{1}{8 \pi^2} \epsilon_\nu^4 \epsilon_d^2
\left[\ln \frac{M_{\rm GUT}}{M_{R_1}} + \ln \frac{M_{\rm GUT}}{M_{R_2}} + \frac{Y^\nu_{33}}{\epd \epn^2}\ln \frac{M_{\rm GUT}}{M_{R_3}}\right] \frac{3 m_0^2 + A_0^2}{m_0^2 + 0.5 M_{1/2}^2}\,. \label{eq:delta23}
\eea
This effect can be dangerously large, only if $\epsilon_\nu\simeq\mathcal{O}(1)$. In this case, we get $(\delta^e_{LL})_{12}\simeq\mathcal{O}(10^{-3})/{\mathcal{R}}$, a value for which the present limit on ${\rm BR}(\mu\to e\gamma)$
already excludes a sizable region of the SUSY parameter space \cite{Masina:2002mv}.
On the other hand, already $\epsilon_\nu\simeq 0.4$ would give a running effect such that
$(\delta^e_{LL})_{12}\simeq\mathcal{O}(10^{-5})/{\mathcal{R}}$, a value which is not very constraining at present, at least in the moderate
$\tan\beta$ regime.

\begin{figure}[t]
\begin{center}
\includegraphics[width=0.45\linewidth]{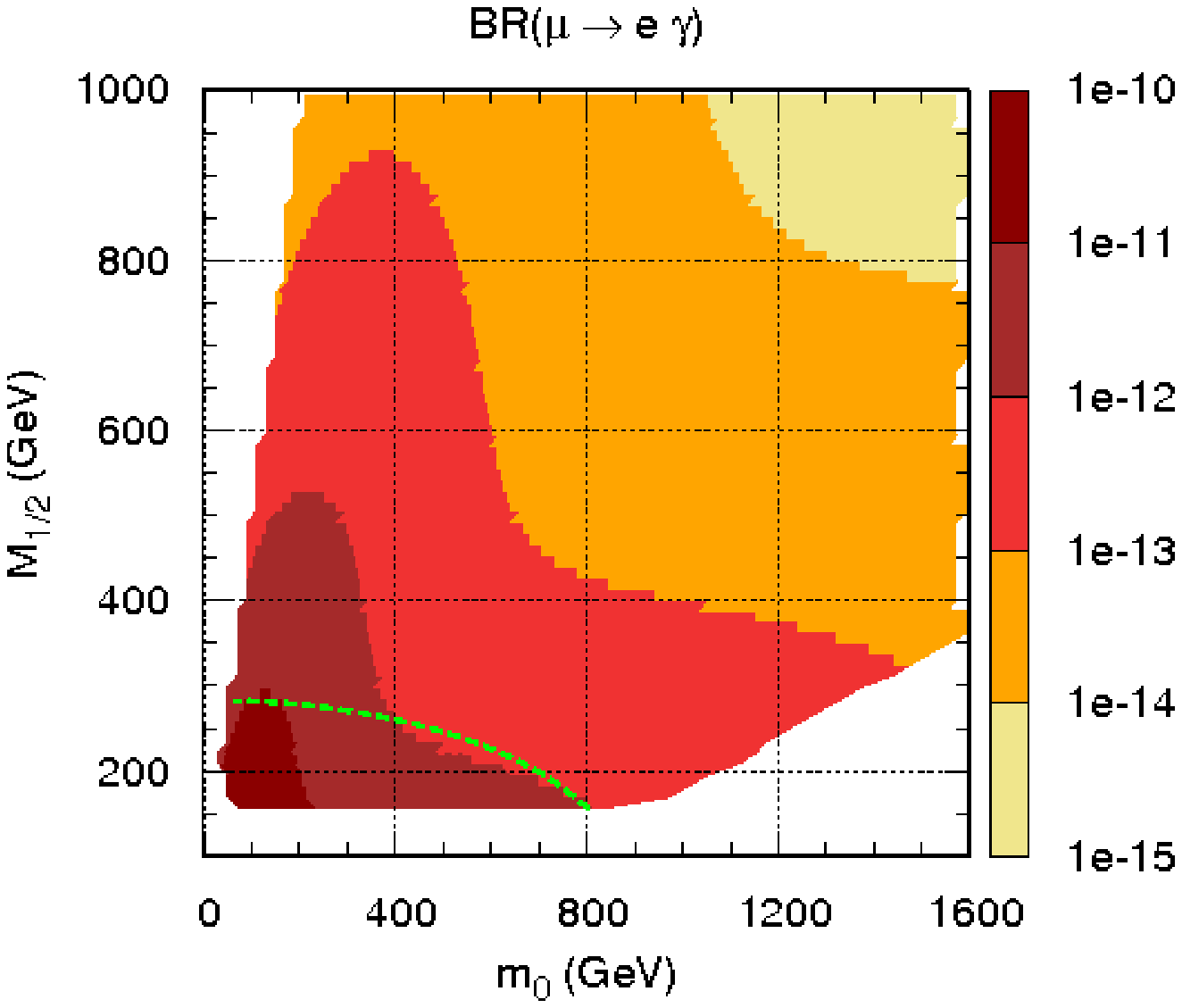}
\hspace{1cm}
\includegraphics[width=0.45\linewidth]{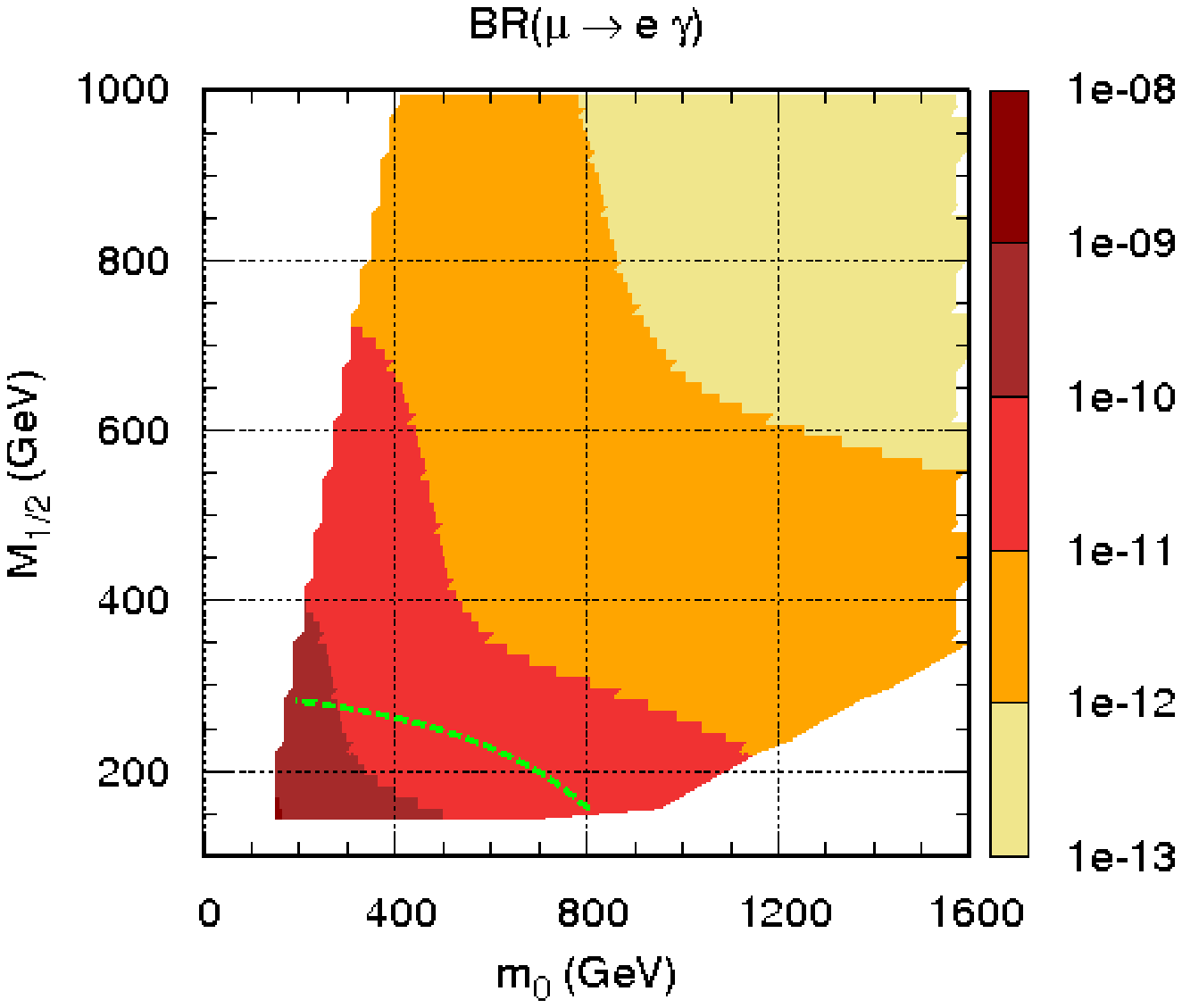}
\caption{Predictions for ${\rm BR}(\mu \to e\gamma)$ in the $(m_0,~M_{1/2})$ plane,
for $\tan\beta=10$ (left) and $\tan\beta=40$ (right), $A_0=0$, $\epsilon_\nu=0.3$. \label{fig:meg}}
\end{center}
\end{figure}
We have performed a numerical evaluation of the RG running and the LFV decay rates. The results are plotted in Fig.~\ref{fig:meg} in the $(m_0,~M_{1/2})$ plane, for $A_0=0$ and $\tan\beta=10$ (left panel) and $\tan\beta=40$ (right panel). In both
cases, we took $\epsilon_\nu=0.3$, a value which can account for the measured baryon asymmetry, as we will see in the
next section. The green dashed-line represents the current LEP bound on the Higgs boson mass (taking into account a theoretical
error of 3 GeV).
The unknown $O(1)$ coefficients in the soft mass matrix of \eq{eq:softSCKM-RR} have been taken to be 1,
therefore variations of the $\delta^e_{RR}$ contribution are possible.
We can see that the moderate $\tan\beta$ regime is practically not constrained by the current experimental
limit ${\rm BR}(\mu\to e\gamma)< 1.2 \times 10^{-11}$, while the final sensitivity ($\simeq 10^{-13}$) of the MEG experiment \cite{meg} will be able to test a large portion of the parameter space. In the large $\tan\beta$ regime the parameter space is already rather constrained and MEG will test it up to SUSY masses well beyond the LHC sensitivity reach.
In case of $A_0\neq0$, we have found that the large contribution from $(\delta^e_{\rm LR})_{12}$ already excludes the parameter space for $m_0 \lesssim 0.8-1 ~{\rm TeV}$ even in the moderate $\tan\beta$ regime, if $A_0/m_0 \simeq 1$. This result is valid up to variations of the unknown $O(1)$ coefficients in the soft matrices $m^2_{\tilde e}$, $A^e$.

Regarding the LFV $\tau$ decays, we have found that the present bound on ${\rm BR}(\mu \to e\gamma)$ excludes
the possibility of observing LFV $\tau$ decays in the foreseeable experiments. In fact, for instance:
$$
{\rm BR}(\tau\to \mu\gamma) \simeq {\mathcal O}(10)\times {\rm BR}(\mu\to e\gamma)\,,
$$
which is a consequence of the fact that MIs in the sector $2-3$ are not much larger than those in the sector $1-2$: indeed, we have  $(\delta^e_{\rm LL})_{23}\sim (\delta^e_{\rm LL})_{12}$ and $(\delta^e_{\rm RR})_{23}\sim (\delta^e_{\rm RR})_{12}/\epsilon_d$,
as we can see from Tab.~\ref{tab:flav_violt_pars} and Eqs.~(\ref{eq:delta12}, \ref{eq:delta23}).

Finally we briefly comment about the flavour violation in the squark sector.
The same expressions of the mass insertions given in Tab. \ref{tab:flav_violt_pars} can be used for the down-quark sector,
since we are not taking into account the $\mathcal{O}(1)$ coefficients that make the differences in these two sectors.
In order to obtain the numerical value of the hadronic mass insertions, we need to take into account a larger factor $\mathcal R$. In fact, for squarks, we typically have:
\be
\mathcal{R}\simeq
\frac{m_0^2 + 6 M_{1/2}^2}{m_0^2}.
\ee
In the up-quark sector, the parameters $(\delta^u_{\rm XX})_{ij}$ are like those of $(\delta^d_{\rm XX})_{ij}$ with the replacement $\epsilon_d\rightarrow \epsilon_u$ and
\be
(\delta^u_{\rm LR})_{12,13} \simeq v \frac{v \tan\beta}{\sqrt{1+\tan\beta^2}}\frac{A_0 \epsilon_u^3}{{\mathcal{R}} m_0^2}\,, ~~
(\delta^u_{\rm LR})_{23}\simeq v \frac{v \tan\beta}{\sqrt{1+\tan\beta^2}} \frac{A_0 \epsilon_u^2}{{\mathcal{R}} m_0^2}\,.
\ee
Bounds of these type have been also analyzed in \cite{Antusch:2007re}. Flavour changing neutral current (FCNC) processes strongly constrain the MIs $\delta^d_{ij}$, especially in the 1-2 sector, while the up sector is at present less constrained \cite{Altmannshofer:2009ne}. From the bounds provided in \cite{Altmannshofer:2009ne}, we see that the model is able to take hadronic FCNC under control.

We would like to remind the reader that the results presented above have a non-trivial dependence on the assumptions we made about the messengers. 
Thus, we can state that a full specification of the messenger sector seems to be unavoidable, in order to improve the predictivity of the $SU(3)$ flavour models.

\section{Constraints from leptogenesis}

For the proper treatment of leptogenesis leading to the observed baryon asymmetry of the Universe, we need to consider the flavour effect (see, for instance, Ref.~\cite{Abada:2006ea} and references therein) in the models under consideration, like it was taken into account for the RHND in \cite{Antusch:2006cw}.

For our calculation of leptogenesis, we will use the approximate
analytic formula derived in Ref.~\cite{Abada:2006ea} taking into
account the arbitrary order-one coefficients and the three
right-handed masses constrained in the section 4.

Since the lightest mass $M_1$, for the normal hierarchy of oscillating neutrinos that we are considering, is in the range of $10^9-10^{12}$ GeV,  we can have the situations of the tau Yukawa  or the muon and tau Yukawa interactions in equilibrium depending on $\tan\beta$.

When $(1+\tan^2\beta)\times10^9 \mbox{ GeV} < M_{I}$,  the tau Yukawa interaction is in equilibrium and thus we use the supersymmetric formula for the baryon asymmetry $Y_B\equiv n_B/s$ normalized by the entropy density $s$ as follows;
\begin{equation}
 Y_B \approx-\frac{10}{31 g_*} \left( \epsilon_{I,2}\, \eta\left( \frac{541}{761} m_{I,2}\right)
 + \epsilon_{I,\tau}\, \eta\left(\frac{494}{761} m_{I,\tau} \right) \right),
\end{equation}
where $g_*=228.75$,   $\epsilon_{I,2} = \epsilon_{I,e} + \epsilon_{I,\mu}$ and
$m_{I,2}=m_{I,e} + m_{I,\mu}$. 
Where we can have $I=1$ or $2$ for a normal or inverted hierarchy ($M_1<M_2$ or $M_1>M_2$) respectively.
  The quantities $\epsilon_{I,\alpha}$, $m_{I,\alpha}$ and the function $\eta$ will be defined  below.

When $  M_{I} < (1+\tan^2\beta)\times10^9 \mbox{ GeV} $,  the muon and
tau Yukawa interaction are in equilibrium and the baryon asymmetry $Y_B$ is given by
\begin{equation}
 Y_B \approx -\frac{10}{31 g_*} \left(
 \epsilon_{I,e} \,\eta\left( \frac{93}{110} m_{I,e}\right)
 + \epsilon_{I,\mu} \, \eta\left( \frac{19}{30} m_{I,\mu}\right)
 + \epsilon_{I,\tau}\, \eta\left( \frac{19}{30} m_{I,\tau}\right)  \right) \,.
\end{equation}
Here we define
\bea
\epsilon_{I,\alpha} &\equiv&
 \frac{1}{8\pi}
\frac{ \sum_{J\neq I}\mathrm{Im}\left[
(y_{\nu}^{\dagger})_{I\alpha}[y_{\nu}^{\dagger}y_{\nu}]_{IJ}
(y_{\nu}^T)_{J\alpha}
\right]
}{
(y_{\nu}^{\dagger} y_{\nu})_{II} }\,
g\left(\frac{M_J^2}{M_I^2}\right) \,, \label{eq:epMSSMaa} \\
\widetilde{m}_{I,\alpha } &\equiv&
(y_{\nu}^{\dagger})_{I \alpha}(y_{\nu})_{\alpha I}\frac{v_{\rm u}^2}{M_I} \,,
\eea
where $g(x) \approx -\frac{3}{\sqrt{x}}$ when $x\gg 1$.
 Finally the so-called wash-out
function $\eta$ can be approximated  \cite{Abada:2006ea} as
\be
 \eta(m) \approx \left( \left(\frac{8.25\times10^{-3} \mbox{ eV}}{m}\right)
                + \left(\frac{m}{ 0.2\times10^{-3}\mbox{ eV}}\right) \right)^{-1} .
\ee

In Fig.~3, we calculate the baryon asymmetry as a function of
$Y^\nu_{11}$ and $\epsilon_\nu$. Here the high (low) $\tan\beta$
region is defined by $1+\tan^2\beta > (<)\; M_1/10^9\mbox{GeV}$.
The region above the horizontal line is allowed as we assumed the
maximal CP violation. For high $\tan\beta$ (the left plot) the
flavour effect becomes important and thus larger $Y^\nu_{11}$
yields larger baryon asymmetry. One can also see that the required
baryon asymmetry can be obtained for $\epsilon_\nu$ larger than
about 0.3.  For low $\tan\beta$ (the right plot) the flavour
effect is suppressed and $\epsilon_\nu \gtrsim 0.3$ is needed as
in the large $\tan\beta$ case.

\begin{figure}[t]
\begin{center}
\includegraphics[width=0.4\linewidth]{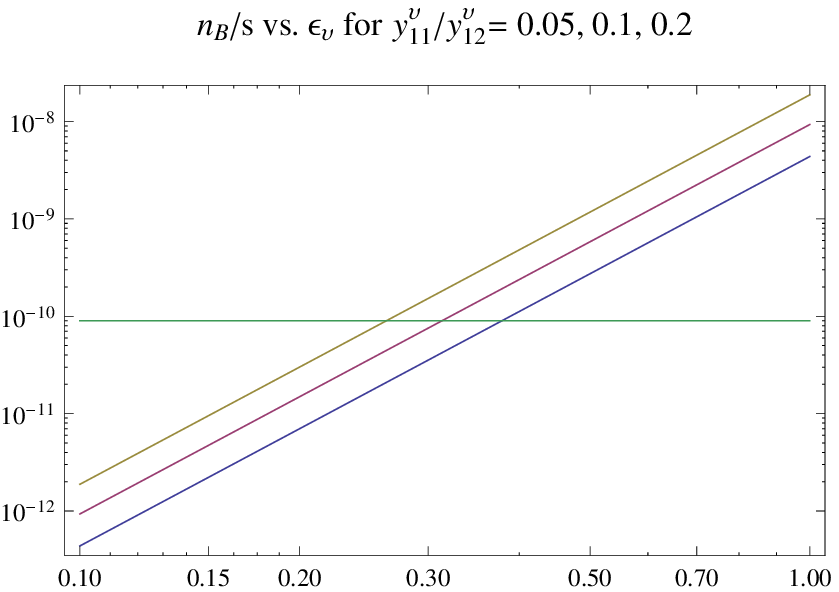}
\includegraphics[width=0.4\linewidth]{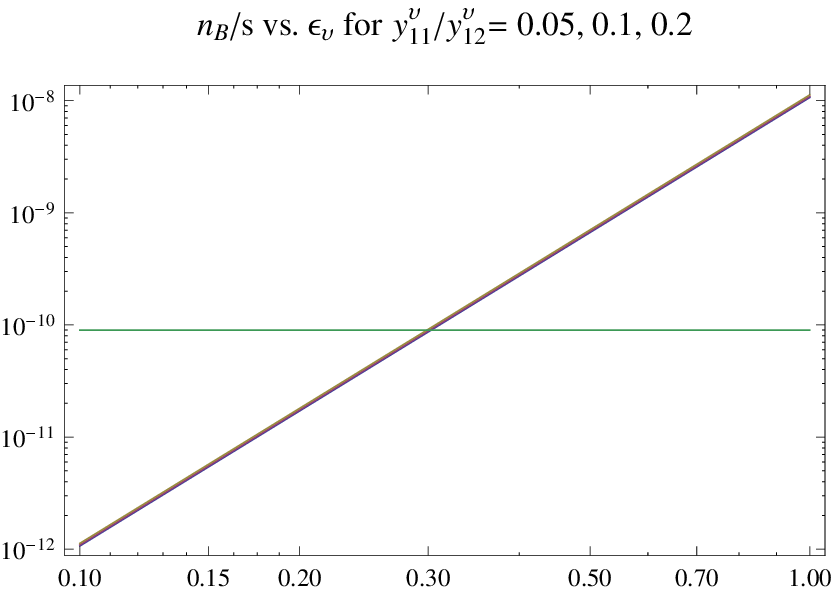}
\caption{ Baryon asymmetry for the normal hierarchy with high
(left figure) and low (right figure) $\tan\beta$. The left figure
shows an enhancement for lager $Y^\nu_{11}$, whereas the curves in
the right figure are insensitive to $Y^\nu_{11}$. \label{YB-NH}} 
\end{center}
\end{figure}

Note that the successful thermal leptogenesis in our framework
requires the right-handed neutrino mass $M_1 \sim 10^{11-12}$ GeV.
These values generally contradict with the standard bigbang
nucleosynthesis due to the decay of the unstable gravitino
\cite{Kawasaki04}. However this gravitino problem can be
circumvented if the gravitno mass is in the range of  $3-100$ TeV
depending on the hadronic branching ratio of the gravitino
\cite{Kawasaki04}.  Another simple way of avoiding the gravitino
problem is to take the axino as the lightest supersymmetric
particle  and the gravitino as the next lightest supersymmetric
particle so that the gravitino decays only to the axino and axion
which has no effect in bigbang nucleosynthesis. In this case, dark
matter can be a composed of the warm axino \cite{Rajagopal} and
the cold axion.

\section{Conclusions}

We have addressed the compatibility of $SU(3)$ family models and the observed value of baryon asymetry through the decay of the lightest right-handed neutrino of a mass about $M_1 \sim 10^{11-12}$ GeV and that at the same time can also satisfy the stringent LFV bounds.

For this purpose we have first updated the fit of the form of the Yukawa matrices that are traditionally embeded in $SU(3)$ models \cite{King:2003rf}-\cite{Calibbi:2009ja}. There is a change of the parameters describing these textures but their form remains valid. Then we have re-stated the conditions under which the $SU(3)$ family symmetry breaking can yield the appropriate quark and lepton (including neutrino) Yukawa couplings. This is done through the VEVs of three flavon fields associated with appropriate heavy masses of only right-handed messengers. These messengers carry non-trivial Pati-Salam (PS) quantum numbers and therefore models of this class can be embedded in a PS model and hence also into a $SO(10)$ GUT. 

While the structure of the right-handed messengers after the breaking of the PS group is similar to that one studied in other $SU(3)$ family models \cite{King:2003rf}-\cite{Calibbi:2009ja} , the structure of the messengers for the right-handed neutrinos is different, essentially because we were interested in raising the scale of the lightest right-handed neutrino to achieve the observed value of baryon asymmetry through its decay.

We therefore have proposed a specific choice of messengers fields allowed by all the symmetries of the model.
Working out all the details of how the canonical normalization of the matter fields with the K\"ahler metric derived from the proposed right-handed messengers, affects the form of the Yukawa couplings and the supersymmetric FV parameters, we have learned that in order to have a real predictive model, the masses of the messenger fields should be explained with further model building ingredients.
 However, we have not constructed a potential from which the particular conditions of messengers could be explained. We would like to point out that indeed for improving the predictivity of $SU(3)$ models, its messenger structure needs to be fully addressed.

  For values of the parameters fitting the measured observables in the quark and lepton sectors, the LFV process $\mu \to e \gamma$ can easily evade the current constraint but can be probed in the MEG experiment for $\epsilon_\nu \gtrsim 0.3$, a value which is consistent with the observed baryon asymmetry.  In the neutrino sector, the lightest right handed neutrino  ($M_1 \sim 10^{11-12}$ GeV) gives the dominant contribution to the neutrino mass matrix.  In the high $\tan\beta$ region the right baryon asymmetry can not be obtained if $Y^\nu_{11} \to 0$ (in the flavour basis) which is the usual requirement when constructing  $SU(3)$ family or $SO(10)$ GUT models in order to reproduce the Gatto-Sartori-Tonin relation \cite{Gatto:1968ss}.
However, non-zero values of $Y^\nu_{11}$ can arise in the canonical basis where the K\"ahler terms are diagonalized and most importantly when considering $Y^\nu$ in the basis where charged leptons are diagonal. It is this last contribution that turns out to be of the required order by the leptogensis processes, which are better understood in this basis.

 In our example in the latter case, this leads to a sizable lepton asymmetry.  For instance, the observed baryon asymmetry can be obtained for $\epsilon_\nu > 0.3$ with $Y^{\nu}_{11}/Y^\nu_{12}=0.1$.  In the low $\tan\beta$ region the size of $Y^\nu_{11}$ is irrelevant for the size of baryon asymmetry and the compatibility with leptogenesis can be obtained also for $\epsilon_\nu \gtrsim 0.3$.

Finally, let us remark that the constraints from leptogenesis imply a rate for $\mu \to e \gamma$ testable by the MEG experiment, at least for a SUSY spectrum in the LHC reach. This is due to the fact that both the CP asymmetry and BR($\mu \to e \gamma$) grow directly with the $Y_\nu$ entries (i.e. with $\epsilon_\nu$ in our case). In case of a negative result of the MEG experiment, this would clearly disfavour this leptogenesis scenario. Other examples of such possible tension between leptogenesis and LFV are provided for instance in \cite{Petcov:2005jh,Chun:2008tw, Calibbi:2009wk}.

\section*{Acknowledgements}
We would like to thank J. Kersten, K. Kadota and S. Antusch for useful conversations, L. V-S and L.C. thank the INFN, The Galileo Galilei Institute for Theoretical Physics (GGI), Firenze, and The Korean Institute for Advanced Study, Seoul, where parts of this work were carried out, for the warm hospitality and the financial support. E.J.C. was supported by Korea Neutrino Research Center through National Research Foundation of Korea Grant (2009-0083526).

\appendix

\section{Fit of the elements of the Yukawa matrices \label{Ap:FIT}}
\subsection{Quark sector}
\begin{table}
\begin{center}
\begin{tabular}{|l l l  l|}
\hline
\multicolumn{4}{c}{Input Values}\\
\hline
Constraints            & Value                 & & Referen.\\
\hline
&  & & \\
$|V_{us}|$       & $0.2246 \pm 0.0012 $   &    &   \\
$|V_{cb}|$       & $(40.59\pm 0.38 \pm 0.58 )\times 10^{-3}$   &   &    \\
$|V_{ub}|$       & $(3.87 \pm 0.09 \pm 0.46 )\times 10^{-3}$   &                   & \cite{ckm_fit_09} \\
$\tan[\gamma] $   & $ 2.45 \pm 0.1 $ & &  \\
$\frac{m_u}{m_c}$ & $0.0021  \pm  0.0002 $    &    & $\dagger$  \\
$\frac{m_c}{m_t}$ & $0.0074  \pm  0.0017 $    &    & $\dagger$     \\
$\frac{m_d}{m_s}$ & $0.0525  \pm  0.0035 $    &    & $\dagger$     \\
$\frac{m_s}{m_b}$ & $0.016  \pm  0.006 $    &    & $\dagger$     \\
$\dagger$ Determined from   &                    &   At $M_Z$ &    \\
$\frac{m_u}{m_d}$    & $0.56 \pm 0.125$  & $0.56 \pm 0.125$      & \\
$\frac{m_c}{m_s}$    & $12.1  \pm 3.1$    & $10.42\pm 3.1 $    & \\
$\frac{m_s}{m_b}$    & $0.025\pm 0.006$  &  $0.016\pm 0.006$    & \\
$Q=m_s/m_d/\sqrt{1-\left(m_u/m_d\right)^2}$                  & $23.0  \pm 2.0$    &       & \\
$m_t$             & $(171.1 \pm 2) $ GeV &   &    \\ \hline
\end{tabular}
\end{center}
\caption{\small Input values for constraints and other related values} \label{tbl:input_par}
\end{table}
\begin{table}
\begin{center}
\begin{tabular}{|l l l l l |}
\hline
\multicolumn{5}{c}{Other values from the CKM fitter}\\
\hline
Parameter            & CKM value   $\pm 1\sigma$ C.L. &  $\pm 2\sigma$ C.L.
& Direct exp. value  $\pm 1\sigma$ C.L. &  $\pm 2\sigma$ C.L.   \\
\hline
$\alpha$ & $90.6^{+3.8}_{-4.2}$    &   $^{+7.5}_{-6.3}$   & $95.6.0^{+3.3}_{-8.8}$
& $^{+5.2}_{-11.8}$  \\
$\beta$  & $21.58^{+0.91}_{-0.81}$ &   $^{+1.8}_{-1.4}$  & $21.07^{+0.90}_{-0.88}$
& $^{+1.8}_{-1.7}$ \\
$\gamma$ & $67.8^{+4.2}_{-3.9}$    &   $^{+6.3}_{-8.0}$   & $70.0^{+27}_{-30}$
& $^{+44}_{-41}$ \\
\hline
\end{tabular}
\end{center}
\caption{\small Relevant information from experiments and from the CKM fitter \cite{ckm_fit_09}. }
\label{tbl:other_ckm_par}
\end{table}
 \begin{table}[ht]
 \begin{center}
 \begin{tabular}{|r|l|l|}
 \hline
 \multicolumn{3}{|c|}{{Quark Fitted Parameters}}\\ \hline
  Parameter & BFP Value & Error  \\
  $|a_{12}^u|$    &   $ 1.33  $ &  -\\
  $a_{22}^d$    &   $ -0.3 $ & $0.002 $ \\
  $a_{22}^u$    &   $ 1.3$ & $0.5$ \\
  $a_{12}^d$    &   $ 0.67$ &  $ 0.0033  $ \\
  $a_{23}^d $   &   $1.52$ &  $0.01$\\
  $\epsilon_u$  &   $0.037$ & $0.0023$ \\
  $\epsilon_d$   &  $0.115$ &  $0.001$  \\ \hline
  \multicolumn{3}{|c|}{{Quark Fixed Parameters}} \\ \hline
$a_{21}^d$  & $ a_{12}^d$ &  \\
$a^d_{13}=a^d_{31} e^{i\Phi_2}$  & $ -1.3 e^{i\Phi_2} $ & \\
$a_{32}^d$  & $ a_{23}^d$ & \\
$a_{21}^u$  & $ |a_{12}^u| $ & \\
$a_{13}^u$  & $  1.0 $ & \\
$\Phi_2$    & $  -1.104$ & \\
$\Phi_1$    & $  -\pi$ & \\
 \hline
 \multicolumn{3}{|c|}{Scaling of observables from $\ME$ to $\MG$}\\\hline
$|V_{cb}|_{\Gt}=$  & $\chi^2 |V_{cb}|_{\Ew} $ & $\chi = 0.7 $\\
$|V_{ub}|_{\Gt}=$  & $\chi^2 |V_{ub}|_{\Ew} $ & \\
$\left|\frac{m_c}{m_t}\right|_{\Gt}=$ & $\chi^4 |\frac{m_c}{m_t}|_{\Ew} $ & \\
$\left|\frac{m_s}{m_b}\right|_{\Gt}=$ & $\chi^4   |\frac{m_s}{m_b}|_{\Ew} $ & \\
\hline
 \end{tabular}
 \end{center}
 \caption{Fitted and fixed parameters for $\tan\beta$ small. Here $a^u_{12}=|a^u_{12}|e^{i\Phi_1}.$  The error in the element $|a_{12}^u|$ is not properly determined by this fit, which reflects a small tension with the assumption $y_b \ll y_t $ .}
 \label{tbl:f_12_13_diffsgn_small_tbeta}
 \end{table}
 \begin{table}[ht]
 \begin{center}
 \begin{tabular}{|r|l|l|}
 \hline
 \multicolumn{3}{|c|}{{Quark Fitted Parameters}}\\ \hline
  Parameter & BFP Value & Error  \\
  $|a_{12}^u|$    &   $ 1.41  $ & $0.05$\\
  $a_{22}^d$    &   $ -0.68$ & $0.16$ \\
  $a_{22}^u$    &   $  1.52 $     & $0.05$ \\
  $a_{12}^d$     &  $  1.30$ &  $0.1$ \\
  $a_{23}^d $   &   $  1.7$   &  $0.05$\\
  $\epsilon_u$    & $  0.04$   & $0.002$ \\
  $\epsilon_d$   &  $  0.13$ &  $0.003$  \\ \hline
  \multicolumn{3}{|c|}{{Quark Fixed Parameters}} \\ \hline
$a_{21}^d$  & $ a_{12}^d$ &  \\
$a^d_{13}=a^d_{31}  e^{i\Phi_2}$  & $ -1.3  e^{i\Phi_2} $ & \\
$a_{32}^d$  & $ a_{23}  $ & \\
$a_{21}^u$  & $ a_{12}^u $ & \\
$a_{13}^u$  & $  1.0 $ & \\
$\Phi_2$    & $  -1.104$ & \\
$\Phi_1$    & $  -\pi$ & \\
 \hline
 \multicolumn{3}{|c|}{Scaling of observables from $\ME$ to $\MG$}\\\hline
$|V_{cb}|_{\Gt}=$  & $\chi |V_{cb}|_{\Ew} $ & $\chi = 0.7 $\\
$|V_{ub}|_{\Gt}=$  & $\chi |V_{ub}|_{\Ew} $ & \\
$\left|\frac{m_c}{m_t}\right|_\Gt =$ & $\chi^3 |\frac{m_c}{m_t}|_{\Ew} $ & \\
$\left|\frac{m_s}{m_b}\right|_\Gt =$ & $\chi   |\frac{m_s}{m_b}|_{\Ew} $ & \\
\hline
 \end{tabular}
 \end{center}
 \caption{Fitted and fixed parameters for $\tan\beta$ large (i.e  $y_b \sim y_t $ ). Here $a^u_{12}=|a^u_{12}|e^{i\Phi_1}.$}
 \label{tbl:f_12_13_diffsgn_large_tbeta}
 \end{table}
The diagonalizing matrices of Yukawa matrices of the form \eq{eq:yuk_o_text}, such that
\bea
\hat{Y}^f=V^f_L Y^f V^{f\dagger}_R,
\label{def:diagz_matxs}
\eea
are to a great approximation given by $V_L^\dagger=U[\ts_{23}]U[\ts_{13}]U[\ts_{12}]$ where $U[\ts_{ij}]_{ij}=s_{ij}=\sin(\theta_{ij})=-U[\ts_{ij}]_{ji}$,  $U_[\ts_{ij}]_{ii}=\tc_{ij}$ and the rest elements are zero in each matrix, where we have omitted the flavour index $f$. Then the angles in each sector are given by $\ts_{12}=a_{12}/(a_{22}-a_{23}a_{32}\epsilon^2) \epsilon $, $\ts_{13}=a_{13}/a_{33} \epsilon^3$, $\ts_{23}=a_{23}/a_{33}\epsilon^2$. We choose only to assign one phase in the $d$ sector, such that $Y^d_{13}=|Y^d_{13}| e^{i \Phi_2}$ and one in the $u$ sector: $Y^u_{12}=|Y_{12}^u| e^{i \Phi_1}$. In this notation just $U[\ts_{13}^d]$ and $U[\ts_{12}^u]$ are complex.

 In terms of the diagonalizing matrices that we use for each sector, the CKM parametrization is given by $V^u_L V^{d\dagger}_L$ but we have to re-phase it before compare it to the standard notation. In this CKM standard notation the observables, \eq{exp:pr_to_fit}, that we such to fix are given to a good approximation by
\bea
V_{us}&=&\left|\frac{|Y^d_{12}|}{Y^d_{22}}-\frac{|Y^u_{12}|}{Y^u_{22}}e^{i\Phi_1}\right|\nonumber\\
V_{ub}&=& \frac{Y^d_{13}}{Y^d_{33}} e^{i\phi_2} - \frac{|Y^u_{12}|}{Y^u_{22}} s_{23}^Q e^{i\Phi_1},\quad s_{23}^Q=\frac{a^d_{23}}{a^d_{33}}\epd^2-\frac{a^u_{23}}{a^u_{33}}\epu^2, \ \delta={\text{Arg}}[V^*_{ub}]\nonumber\\
V_{td}&=&-\frac{Y^d_{13}}{Y^d_{33}} e^{-i\phi_2} + s^Q_{23} \frac{Y^u_{12}}{Y^u_{22}} e^{-i\Phi_1}  V_{us}.
\label{rel:ckm_this_par}
\eea
Since we have more parameters to fit than observables we need to fix some of the parameters in the Yukawa matrices, to this end we choose to fit the parameters which are more sensitive to the observables we have chosen. We can see that the parameters determining the right-handed diagonalizing matrices of the Yukawa matrices, or sub-dominant contributions of the left-handed diagonalizing matrices, are not that crucial and therefore we scan a set of fixed values, perform a fit to the other parameters which have not been fixed and then choose the best fit of the scan on the fixed parameters. Although the parameter $a^d_{13}$ is relevant for the observables we choose to fit, it is easier to fix it, instead of fitting it because of the phase involved in the term $Y^d_{13}$. The fit is done with the help of MINUIT and the ROOT environment \cite{root_cern} and with the exact expressions of the elements $V_{ij}$ in terms of the mixing angles, not with the approximation of \eq{rel:ckm_this_par}.
We want to emphasize here that it is easy to see why once fixing the absolute value of the CKM angles in \eq{exp:pr_to_fit} and assuming an unitary CKM matrix, which is consistent in our set up of three families, then the angles $\beta$ and $\alpha$ are also fixed and are in agreement with the experimental observations. Using the standard CKM parametrization we see that $\gamma\approx \delta$,  $\beta\approx Arg[-V_{td}]$ and obviously then  $\alpha\approx \delta+ \beta$. Note that this is consistent with what it was used in \cite{Roberts:2001zy} ($\delta=\pi\pm \Phi_1 -\beta$), because we satisfy $\delta=\pi+\Phi_1-\beta=\pi-\beta$.
 The element $V_{td}$ is given to a good approximation by $|V_{us}| |V_{cb}| - V^*_{ub}$ then fixing each of these values with inputs given in Table \ref{tbl:input_par} we ensure to have $\beta \approx 22^o$. Using the expressions in \ref{rel:ckm_this_par} and the values obtained from the fit, in Tables \ref{tbl:f_12_13_diffsgn_large_tbeta} and \ref{tbl:f_12_13_diffsgn_small_tbeta}, one can also easily check that the unitary angles are given to a good agreement within the allowed experimental range. The Yukawa matrices from \eq{eq:yuk_o_text} should be identified with a particular family model at the scale at which its family group it is broken, we usually think for simplicity that this scale can be identified to the GUT scale. In this case one still need to take into account the running of this texture from $\MG$ to $\ME$. Doing this numerically involves a good effort but we can take a shortcut by choosing to fit parameters from which we know that the exact details of the supersymmetric spectra are not that relevant, just as it was done in \cite{Roberts:2001zy}. We know that for precision analysis the effects of the supersymmetric spectra is relevant \cite{Olive:2008vv,Antusch:2008tf} however we can follow \cite{Olechowski:1990bh} and use the ratios of the quark masses and the CKM elements that we choose to fit, \eq{exp:pr_to_fit}, because we can easily estimate how do they scale from $\ME$ up to $\MG$ as long as we start with universal supersymmetric conditions. In this case there are two different limiting cases, one for which $y_b \ll y_t $ that is $\tan\beta$ small (Table \ref{tbl:f_12_13_diffsgn_small_tbeta}) and the other  for which $y_b \sim y_t $ that is $\tan\beta$ large (Table \ref{tbl:f_12_13_diffsgn_large_tbeta}).

%
%
\subsection{Neutrino sector}
We have performed a simple fitting analysis to the parameters of the neutrino sector described in $\S$ \ref{sec:leptmix}. The results of it are presented in Table \ref{tbl:neut_par}.
\begin{table}[t]
\begin{center}
\begin{tabular}{|l |l |}
\hline
\multicolumn{2}{|c|}{Neutrino Fitted Parameters}\\
\hline
$a^\nu_{21}$       & 0.9    \\
$a^\nu_{31}$       & 1.21    \\
$a^\nu_{22}$       & $0.63$   \\
$a^\nu_{23}$       & $-0.5$   \\
$a^\nu_{23}$       & $-1$\\
\hline
\multicolumn{2}{|c|}{Neutrino Fixed Parameters}\\
\hline
$a^\nu_{11}$       & $0.7$  \\
$a^\nu_{12}$       & 1    \\
$a^\nu_{13}$       & 1    \\
$\epsilon_\nu$    & $0.3$\\
\hline
\multicolumn{2}{|c|}{O u t p u t s}\\
\hline
\multicolumn{2}{|c|}{Angles }\\
\hline
$\tan[\theta_{13}]$ & $0.03$\\
$\tan[\theta_{12}]$ & $0.68$\\
$\tan[\theta_{23}]$ & $0.93$\\
\hline
\multicolumn{2}{|c|}{Masses }\\
\hline
$m_1=0.0008$ eV & $M_1=1.9\times 10^{11}$ GeV\\
$m_2=0.0087$ eV & $M_2=1.06\times 10^{12}$ GeV \\
$m_3=0.05$ eV&  $M_3= \times 10^{13}$ GeV\\
\hline
\end{tabular}
\end{center}
\caption{\small  Values for parameters of the neutrino sector, compatible at the 1 sigma for all the fitted values of the global fit, except for $\theta_{13}$ which is compatible at 2 sigma, of the reference \cite{GonzalezGarcia:2010er}. Of course for  $\theta_{13}$ the relevant quantity to compare, it is the upper limit. The value of $a^\nu_{11}$ has not been fitted but constrained using the relation \eq{eq:rel_anu11}. For the fitted values a conservative $5\%$ 1 sigma error can be assumed.}\label{tbl:neut_par}
\end{table}

\section{Superpotential before decoupling of heavy fields and K\"ahler metric}

If we do not introduce left-handed messengers (or we set $M_L\gg M_R$), the renormalizable superpotential above the messengers scale reads:
\bea
W &=& \psi  H  X_R^3 + \overline{X}_R^3 \left(g_a\, \overline{\phi}_{123} \chi^1_R  + g_a^\prime\, \overline{\phi}_{3} \chi^{1 \,\prime}_R
+g_a^{\prime\prime}\, \overline{\phi}_{23} \chi^{1\,\prime\prime}_R \right) + \nonumber \\
&& \left(g_b\, \overline{X}^1_R \overline{\phi}_{23} + g_b^\prime\, \overline{X}^{1\,\prime}_R \overline{\phi}_{3}  +
g_b^{\prime\prime}\, \overline{X}^{1\,\prime\prime}_R \overline{\phi}_{123}\right) \psi^c + g_c\,\overline{X}^{1\,\prime\prime}_R \hat{H}
\chi^{1}_R
\label{eq:renormW}
\eea
The $\mathcal{O}(1)$ coefficients appearing in the Yukawa matrices in Eq.~(\ref{eq:yuk_o_as_int_insu3}), which is written in the flavour basis, in terms of the fundamental coefficients are
\bea
g_{12}=-g_{13}=-g_ag_b, &~~& g_{21}=-g_{31}=-g_a^{\prime\prime}g_b^{\prime\prime},\nonumber \\
g_{22}= -g_{23}= -g_{32} = g_a^{\prime\prime}g_b g_c , &~~& g_{22}^{\prime}= g_ag_b+g_a^{\prime\prime}g_b^{\prime\prime},\nonumber\\
g_{23}^\prime = -g_{32}^\prime & = & -g_ag_b+g_a^{\prime\prime}g_b^{\prime\prime}, \label{eq:coeff_fud_op}
\eea
and $\epsilon_e = \epsilon_d$, $\mathcal{Y}^u=-2/3$, $\mathcal{Y}^d=1/3$, $\mathcal{Y}^d=1$.

The K\"ahler metric involving the flavon fields that we consider and that it is compatible
with the superpotential of \eq{eq:renormW} it is the following:
\bea
K_{\psi \psi^\dag}= \mathbf{I}, ~~
K_{\psi^c \psi^{c\,\dag}}\simeq \mathbf{I}+
\left(
\begin{array}{ccc}
g_b^{\prime\prime\,2} (\epsilon_d \epsilon_f)^2 & g_b^{\prime\prime\,2} (\epsilon_d \epsilon_f)^2 & g_b^{\prime\prime\,2} (\epsilon_d \epsilon_f)^2 \\
g_b^{\prime\prime\,2} (\epsilon_d \epsilon_f)^2 & g_b^2 \epsilon_f^2 & g_b^2 \epsilon_f^2 \\
g_b^{\prime\prime\,2} (\epsilon_d \epsilon_f)^2 & g_b^2 \epsilon_f^2 &  g_b^{\prime\,2} \frac{a_f^2}{M^2_{\chi^{1\prime}_R}}
\end{array}
\right),
\label{eq:non_dg_kahler}
\eea
where, in the 2-3 sector, we neglected the subdominant terms $\propto |\overline{\phi}_{123}|^2$  and in $K_{\psi_3 \psi^\dag_3}$ also a contribution from $\propto|\overline{\phi}_{23}|^2$, in agreement with the hierarchy of \eq{eq:vevs}.
\providecommand{\bysame}{\leavevmode\hbox to3em{\hrulefill}\thinspace}

\end{document}